# Dynamics of Silent Universes

Marco Bruni

*Astronomy Unit, School of Mathematical Sciences,*

*Queen Mary & Westfield College, Mile End Road, E1 4NS London, U.K.*

*Dipartimento di Astronomia, Università di Trieste, via Tiepolo 11, 34131 Trieste, Italy*

*SISSA, via Beirut 2–4, Trieste 34013, Italy*

Sabino Matarrese & Ornella Pantano

*Dipartimento di Fisica "Galileo Galilei",*

*Università di Padova, via Marzolo 8, 35131 Padova, Italy*

June 25, 1994

**Abstract**

We investigate the local non–linear dynamics of irrotational dust with vanishing magnetic part of the Weyl tensor, $H_{ab}$. Once coded in the initial conditions, this dynamical restriction is respected by the relativistic evolution equations. Thus, the outcome of the latter are *exact solutions* for special initial conditions with $H_{ab} = 0$, but with no symmetries: they describe inhomogeneous triaxial dynamics generalizing that of a fluid element in a Tolman–Bondi, Kantowski–Sachs or Szekeres geometry. A subset of these solutions may be seen as (special) perturbations of Friedmann models, in the sense that there are trajectories in phase–space that pass arbitrarily close to the isotropic ones. We find that the final fate of ever–expanding configurations is a spherical void, locally corresponding to a Milne universe. For collapsing configurations we find a whole family of triaxial attractors, with vanishing local density parameter $\Omega$. These attractors locally correspond to Kasner vacuum solutions: there is a single physical configuration collapsing to a degenerate *pancake*, while the generic configuration collapses to a triaxial *spindle* singularity. These *silent universe* models may provide a fair representation of the universe on super horizon scales. Moreover, one might conjecture that the non–local information carried by $H_{ab}$ becomes negligible during the late highly non–linear stages of collapse, so that the attractors we find may give all of the relevant expansion or collapse configurations of irrotational dust.

# 1 Introduction

A good deal of the work of theoretical physicists is spent in constructing models for natural phenomena out of their theories, this process perhaps eventually being terminated by an experimentalist colleague. As a rule, in this practical process some approximation is taken in order to reduce the problem to a tractable one, the hope being that the approximation considered is sufficiently reasonable that the results produced are still useful, either because they directly give a sufficiently good description of some phenomena or – this goal not being reached – because through them we nevertheless gain some clue to the next possible steps we can take to set up a better model.

Roughly speaking, we may divide possible approximations in two broad and not necessarily disjoint classes.[1]

The first one we may term the class of *exact approximations*, where by the deliberate use of this contradictory definition we want to indicate all those exact solutions of the theory we are using that can be derived under some *special assumptions* regarding, for example, the matter content and/or the boundary conditions. The second class is that of *approximations for general data*: in this we include all those truly approximate solutions of the equations of the theory which can be derived by making some ansatz under which the equations noticeably simplify, but still accept generic boundary conditions.

The search for approximate solutions in these two classes outline two strategies: it seems to us that these are equally important and therefore complementary, especially in dealing with non linear problems such as those posed by Newtonian and relativistic gravity. In looking for exact approximations we often find solutions with unexpected behaviours, revealing aspects of the full non linear dynamics, while approximations for general data give us a clue to what can be expected under general and reasonable circumstances, but only within the range of validity of the underlying assumptions. Examples in the first class are spherical or axisymmetric solutions of general relativity with matter, often showing singular behaviour that then deserves further study. In the second class we have Newtonian and relativistic cosmological perturbations, and the Zel'dovich approximation that reasonably describes pancaking in the first stage of the non–linear regime of Newtonian gravity.

Recently, cosmologists's attention has been drawn on relativistic irrotational dust with vanishing magnetic Weyl tensor, $H_{ab} = 0$ (Matarrese, Pantano, & Saez 1993, Croudace et al. 1994, Bertschinger & Jain 1994). Having in mind to follow cosmological perturbations in the non–linear regime in the matter dominated era, the assumption of dust is the simplest description one can take for the most interesting case of collisionless matter. The irrotational assumption is well justified if we take a broad view, i.e. if we think of the resulting description as valid for not too small scales. Then, the *kinematical restriction* $\omega_{ab} = 0$ is an exact approximation, as it is well known that an initially irrotational flow stays irrotational, in the absence of dissipative effects. Moreover, according to the inflationary paradigm for the generation of perturbations, no initial vector modes are present at the onset of cosmological structure formation.

[1] We do not want to give here a formal classification of possible approximations, we merely introduce these two classes for the sake of the following discussion; in particular we do not claim that these two classes are also exhaustive. For an extended and somehow related discussion, see (Ellis 1993).

The assumption of vanishing magnetic Weyl tensor is the most controversial one. The question that was posed and left open by Matarrese, Pantano, & Saez (1993) was if the chosen setting $H_{ab}=0$ was sufficiently wide to accept generic initial data for irrotational dust. More recent works (Matarrese, Pantano, & Saez 1994a,b; Bertschinger & Hamilton 1994, Kofman & Pogosyan 1994) have indeed proven that generic purely scalar perturbations of a Friedmann–Robertson–Walker (FRW hereafter) universe, although giving $H_{ab}=0$ at first order (Goode 1989; Bruni, Dunsby & Ellis 1992), give rise to a non vanishing magnetic Weyl tensor at second order, an effect that we may term *tidal induction.*

However, if we want to insert the assumption $H_{ab}=0$ (having already assumed $\omega_{ab}=0$) in one of the two classes above, then this is an exact approximation. This was implicitly assumed by Matarrese, Pantano, & Saez (1993) and was in fact proven by Barnes & Rowlingson (1989) in an earlier paper for the more general case of a perfect fluid. In other words, Einstein equations admit exact solutions for an irrotational perfect fluid with vanishing magnetic Weyl tensor, $\omega_{ab}=H_{ab}=0$. Some of these spacetimes are completely general from the point of view of the algebraic classification of the Weyl tensor, i.e. they are of Petrov type I, and others belong to the degenerate type D or O. The latter are conformally flat and all known; type D spacetimes are those for which the gravitational field is purely Coulombian (Szekeres 1965), e.g. as in the case of Scharzschild or Kerr; type I may be characterized by a superposition (because the geodesic deviation equation is *linear* in the curvature) of purely Coulombian and transverse fields (Szekeres 1965). As shown by Barnes & Rowlingson (1989), the dust type D spacetimes with $H_{ab}=0$ are known explicitly, and due to Szekeres (1975a): in the following we will refer to them as Szekeres solutions.

The assumption $H_{ab}=0$ is a *dynamical restriction* (because is a constraint on the tidal field) that has two implications: *i)* the first comes from one of the usual constraint equations (the $H_{ab}$ constraint), which restricts the spatial distribution of the shear $\sigma_{ab}$ (because we also have $\omega_{ab}=0$); *ii)* the second one comes from *an additional constraint equation* that arises when we impose $\dot{H}_{ab}=0$, which restricts the spatial variation of the electric tidal field $E_{ab}$.

The restriction $H_{ab}=0$ may also be regarded as a sort of gauge choice on the curvature (rather than on the metric as it is usually considered in general relativity) that perhaps could be dubbed "no induction" or "no radiation", as this, for example, implies flatness if imposed on an exact vacuum solution (Petrov type N) for a gravitational wave (Bruni & van der Elst 1994); a weaker condition, i.e. the vanishing of $H_{ab}$ at infinity, was shown to be physically equivalent to a certain boundary condition in supergravity (Hawking 1982, 1983).

The interesting feature of the $\omega_{ab}=H_{ab}=0$ perfect fluid spacetimes is that there is an orthonormal tetrad associated with the matter 4–velocity $u^a$ which is the simultaneous eigenframe for the shear $\sigma_{ab}$ of the matter flow and for the electric Weyl tensor $E_{ab}$: in all but two special cases (Barnes & Rowlingson 1989) the vectors of this tetrad are hypersurface orthogonal, so that a coordinate system exists in which the metric $g_{ab}$, $\sigma_{ab}$ and $E_{ab}$ are all diagonal. If, in addition to this, the flow is taken to be geodesic, $\dot{u}^a=0$ (as it is the case for dust), the evolution equations simplify a great deal (Matarrese, Pantano, & Saez 1993), reducing to only six ordinary differential equations. Thus, assuming the constraint equations are satisfied by the initial data, the following evolution of each fluid element is no more influenced by the environment or, in other words, it proceeds as that of a separate universe, which therefore may be dubbed *silent universe* (Matarrese, Pantano, & Saez 1994a,b).

This will be exactly the point of view taken here: assuming the constraint equations are satisfied, we will investigate the local dynamics of silent models in complete generality, mainly using the theory of dynamical systems (e.g. Arrowsmith & Place 1982, Arnol'd 1992) but always comparing the outcome of the latter with numerical results. In particular, we will not need to assume that the initial conditions are necessarily those arising as linear perturbations of a FRW universe, although a subset of such initial conditions is also a subset of the initial conditions that are accepted by the equations for silent universes (more precisely, there is a subset of perturbed FRW initial conditions that satisfies the constraint equations for the $p = \omega_{ab} = H_{ab} = 0$ case *exactly*, so that these initial conditions are also a subset of all the possible initial conditions that satisfy the same constraints). Consistently, we will use a set of covariantly defined variables, making no reference to a background FRW model. The perturbative point of view can always be recovered in the end, although in general, in doing this, there is a problem of gauge choice, basically arising from the fact that in general the initial singularity does not occur at the same time in inhomogeneous models, as it happens for FRW universes.

The formalism we use will be briefly reviewed in Section 2. In Section 3 we will show how the introduction of a convenient set of five dimensionless variables and a new time variable decouples their evolution equations from that for the expansion $\Theta$ (the Raychaudhuri equation). After a discussion of some general properties of this novel set of five equations, we will analyze in Section 4 the subcase of Szekeres models, arising from the simultaneous degeneracy of the shear and electric tidal field eigenvalues. Because of this degeneracy, the Szekeres fluid element may be said locally axisymmetric, although these models admit no Killing vectors. These models have been extensively discussed in the literature (e.g. Kramer et al. 1980, and reference therein), and retain many of the features of the general case, while only three of our variables are needed to discuss their dynamics. Thus, the phase–space for these models is 3–dimensional, and can be visualized, revealing general properties that also hold in the general 5–dimensional case. In particular, our approach will immediately reveal how there are two attracting stationary points for collapsing Szekeres fluid elements, one corresponding to a pancake singularity and the other to a spindle singularity, while the final fate of an ever–expanding patch of these spacetimes is to fall into an attracting point representing a spherical void (a Milne universe). In Section 5 we will come back to the general triaxial case, and we will show that silent models in the collapse phase admit an attracting set which is a closed curve in the 5–dimensional phase–space, corresponding to the Kasner (vacuum Bianchi I) solutions of general relativity. As it is well known (e.g. Zel'dovich & Novikov 1983, Stephani 1990), there is a single pancake in this set, while the generic case is that of the cigar or spindle singularity. The pancake is the same as that of the Szekeres models, i.e. is locally axisymmetric, but the generic spindle is triaxial. Thus, the final fate of a generic silent fluid element that stops expanding is to collapse to a triaxial spindle singularity, although for perturbed FRW initial conditions the final configuration could be almost locally axisymmetric (Bertschinger & Jain 1994). As it is the case for simple homogeneous models (Zel'dovich & Novikov 1983), the final stage of collapse (or the generic initial singularity) is dominated by the curvature, and the matter is unimportant, as shown by the vanishing of the local density parameter, $\Omega$. Again, the attracting point of ever–expanding configurations is that representing a Milne universe.

Finally, we will give the asymptotic behaviour of trajectories around the stationary points of the phase–space of silent universes. This asymptotic behaviour should be seen as the behaviour of a perturbation around the background solution given by the stationary point itself. This analysis will further clarify things like the expansion away from a flat FRW model, the expansion toward a spherical void configuration, and the collapse to a pancake or a spindle.

Conclusions are drawn in Section 6, where we also make some conjectures regarding a number of possible cosmological applications of silent universe models.

Throughout this paper we use units $c = 8\pi G = 1$; our signature is $(-,+,+,+)$.

# 2 Relativistic dynamics

In this section we will briefly summarize a formulation of the relativistic dynamics of collisionless matter with zero velocity dispersion (dust) in terms of covariant variables that represent observable kinematical and dynamical (curvature) quantities, focusing on the case of irrotational dust with vanishing magnetic Weyl tensor: $p = \omega_{ab} = H_{ab} = 0$. We will also give the set of evolution equations for these variables as derived by Matarrese, Pantano, & Saez (1993) (see also Barnes & Rowlingson 1989). These equations are a specialization to the case $p = \omega_{ab} = H_{ab} = 0$ of those presented by Ellis (1971), where a comprehensive presentation to the approach followed here can be found[2].

## 2.1 Hydrodynamical and gravitational field variables

Let us consider a perfect fluid with four–velocity $u^a$, normalized to $u^a u_a = -1$. At each spacetime point we may define a projection tensor into the rest space of an observer moving with the same four velocity $u^a$ (comoving observer): $h^{ab} \equiv g^{ab} + u^a u^b$, with $h_{ab} u^a = 0$. With $u^a$ and $h^{ab}$ the covariant derivative of any tensorial quantity can be split into a time derivative and a spatial derivative. In particular, the spatial part of the derivative $u_{a;b}$ of $u^a$ is given by $v_{ab} \equiv h_a{}^c h_b{}^d u_{c;d}$, with $v_{ab} u^b = 0$, while its time part is the acceleration $\dot{u}^a \equiv u^a{}_{;b} u^b$, which is a space–like vector, $\dot{u}^a u_a = 0$. An overdot denotes convective differentiation with respect to proper time $t$ of comoving observers (for a general $n$–tensor $A$, $\dot{A}_{a_1,a_2 \ldots a_n} \equiv A_{a_1,a_2,\ldots a_n;b} u^b$). It is standard to split the tensor $v_{ab}$ into its trace $\Theta \equiv v^a{}_a$, its symmetric trace–free part $\sigma_{ab} \equiv v_{(ab)} - \frac{1}{3} h_{ab} \Theta$, and its skew symmetric part $\omega_{ab} \equiv v_{[ab]}$ (the symbol $_{[..]}$ stands for skewsymmetrization, $_{(..)}$ for symmetrization). These are *kinematical quantities*, as they determine the relative velocity of neighboring fluid elements. In particular, given a fluid sphere at an initial time, the expansion scalar $\Theta$ gives its average volume expansion ($\Theta > 0$) or contraction ($\Theta < 0$), the shear tensor $\sigma_{ab}$ gives its deformation at fixed volume into an ellipsoid, and the vorticity tensor $\omega_{ab}$ gives its rotation with respect to a locally inertial frame.

It is possible to give an alternative formulation of general relativity (Ellis 1971) using as dynamical variables $\rho$, $\Theta$, $\sigma_{ab}$, $\omega_{ab}$, and the *dynamical quantities* (in that they determine the relative acceleration of fluid elements (Szekeres 1965, Hawking & Ellis 1973) $E_{ab} \equiv C_{acbd}u^c u^d$ and $H_{ab} = \frac{1}{2}\eta_{ac}{}^{gh}C_{ghbd}u^c u^d$, where $C_{abcd}$ is the Weyl tensor, i.e. the trace–free part of the curvature. The symmetric trace–free tensors $E_{ab}$ and $H_{ab}$ are usually named the electric and magnetic part of the Weyl tensor, or simply the electric and magnetic tidal field; they are also flow–orthogonal, $E_{ab}u^b = H_{ab}u^b = 0$. The electric tidal field $E_{ab}$ has a straightforward Newtonian analogue ($E_{\alpha\beta} = \phi_{,\alpha\beta} - \frac{1}{3}\delta_{\alpha\beta}\nabla^2\phi$), while there is no counterpart of $H_{ab}$ having an independent dynamical role in Newtonian theory (Ellis 1971, Kofman & Pogosyan 1994).

[2]Recently the original 1961 review article by Ehlers has been translated into English by G. F. R. Ellis and P. K. S. Dunsby (Ehlers 1993).

In this formulation, the Einstein equations determine a local algebraic relation between the "trace" part of the curvature $R_{ab}$ (the Ricci tensor) and the matter content, as described by the energy momentum tensor $T_{ab}$: $R_{ab} = (T_{ab} - \frac{1}{2}g_{ab}T)$. The role of field equations is then played by the Ricci identities for $u^a$, i.e. $u_{a;d;c} - u_{a;c;d} = R_{abcd}u^b$, from which the evolution equations for the kinematical quantities are derived (as well as a set of constraint equations that must be satisfied by the initial data), and by the Bianchi identities in the form $C^{abcd}{}_{;d} = R^{c[a;b]} - \frac{1}{6}g^{c[a}R^{;b]}$. Thus, the Ricci part of the curvature is *locally* determined by matter through Einstein equations, and using these to substitute for $R_{ab}$ in terms of $T_{ab}$ into the Bianchi identities one has a set of equations in which the trace–free Weyl part of the curvature is determined *non–locally* by matter. Splitting the Weyl tensor $C_{abcd}$ into the electric and magnetic parts $E_{ab}$ and $H_{ab}$ the Bianchi identities become four equations that can be said Maxwell–like, as they resemble Maxwell equations. As usual, the energy and momentum conservation equations $T^{ab}{}_{;b} = 0$ follow from the contracted Bianchi identities.

Here we are interested in a perfect fluid with vanishing pressure, $p = 0$, and with vanishing vorticity, $\omega_{ab} = 0$. Two immediate standard results follow from these assumptions: *i)* since $p = 0$, the acceleration vanishes as a consequence of the momentum conservation equation, so that the fluid flow is geodesic; *ii)* since $\omega_{ab} = 0$, the fluid flow is hypersurface orthogonal, i.e. there exist space–like hypersurfaces of which $u^a$ is the normal vector; one can use spatial coordinates on these surfaces, and proper time along flow lines as coordinate time, thus defining a comoving synchronous coordinate system.

However, one can also use a *frame* rather than a coordinate description, i.e. define components of tensors over a tetrad of vectors. In particular, we can use the orthonormal tetrad $\{u^a, e^a_\alpha\}$, $u_a e^a_\alpha = 0$, $e^a_\alpha e^\beta_a = \delta^\alpha_\beta$ ($\alpha$, $\beta = 1, 2, 3$; here greek indices label these vectors) so that the tetrad components of tensors are *scalars* (e.g. Stephani 1990) that are actually measured by the comoving observers, as for example the matter density $\rho \equiv T_{ab}u^a u^b$.

At this point we can make our third fundamental assumption, i.e. we impose the vanishing of the magnetic tidal field $H_{ab} = 0$. As it was shown by Barnes & Rowlingson (1989) it follows from the field equations (the $divH$ equation) that, with this sort of gauge choice on the curvature, or dynamical restriction, the eigenframe of the shear $\sigma_{ab}$ and of the electric tidal field $E_{ab}$ coincide. Thus, taking the $e^a_\alpha$ aligned with the eigenvectors of $\sigma_{ab}$ and $E_{ab}$, these two quantities may be written as

$$E_{ab} = \sum_{\alpha=1}^{3} E_\alpha e_{\alpha a} e_{\alpha b} \ , \quad \sigma_{ab} = \sum_{\alpha=1}^{3} \sigma_\alpha e_{\alpha a} e_{\alpha b} \ , \tag{1}$$

with

$$\sum_{\alpha=1}^{3} E_\alpha = \sum_{\alpha=1}^{3} \sigma_\alpha = 0 \ , \quad \sigma^2 \equiv \frac{1}{2}\sigma_{ab}\sigma^{ab} = \frac{1}{2}\sum_{\alpha=1}^{3} {\sigma_\alpha}^2 \ , \quad E^2 \equiv \frac{1}{2}E_{ab}E^{ab} = \frac{1}{2}\sum_{\alpha=1}^{3} {E_\alpha}^2 \ , \tag{2}$$

where $\sigma$ and $E$ are the magnitudes of the shear and electric tidal field. It was also proven by Barnes & Rowlingson (1989) that (in all but two special cases) the spatial tetrad vectors $e^a_\alpha$ are also hypersurface orthogonal, so that a coordinate basis exists in which the metric is also diagonal together with $\sigma_{ab}$ and $E_{ab}$. In this case we have

$$u_a = -\delta^4_a \ , \quad e_{\alpha a} = \ell_\alpha \delta^\alpha_a \ , \tag{3}$$

and the metric may be written as

$$ds^2 = -dt^2 + \sum_{\alpha=1}^{3} {\ell_\alpha}^2(\vec{x}, t)(dx^\alpha)^2 \ , \tag{4}$$

where the $\ell_\alpha$'s give the scaling of lengths in the three directions $e^a_\alpha$. Then, a local average scale factor $\ell$ may be defined, so that the mean expansion rate and the expansion rates in the three directions $e^a_\alpha$ are given by

$$\frac{\dot{\ell}}{\ell} = \frac{1}{3}\Theta \ , \quad \frac{\dot{\ell}_\alpha}{\ell_\alpha} = \sigma_\alpha + \frac{1}{3}\Theta \ , \tag{5}$$

that define the $\ell_\alpha$'s up to a factor which is constant along each flow line. Thus, $\ell$ is the geometric mean of the directional scale factors $\ell_\alpha$, while $\frac{1}{3}\Theta$ is the average of their expansion rates:

$$\ell = \prod_{\alpha=1}^{3} {\ell_\alpha}^{\frac{1}{3}} \ , \quad \frac{\dot{\ell}}{\ell} = \frac{1}{3}\sum_{\alpha=1}^{3} \frac{\dot{\ell}_\alpha}{\ell_\alpha} \ . \tag{6}$$

With these definitions, we can better see the effect of the shear $\sigma_{ab}$. An initially spherical fluid element tends to a flattened configuration if it has two non–negative shear eigenvalues and a strictly negative one, which corresponds to having one direction whose expansion rate is lower than the mean local expansion rate $\frac{1}{3}\Theta$. It tends to an elongated configuration if it has two negative shear eigenvalues and a positive one; in this case two directions have expansion rates lower than the local average. When two of the shear eigenvalues are equal, i.e. the shear is degenerate, the pancake–like configuration is an oblate spheroid, while the elongated one is a prolate spheroid.

Finally, following a standard terminology (MacCallum 1973, Goode & Wainwright 1982), we define singularities as: *i) point–like* if all three $\ell_\alpha \to 0$; *ii) cigar* (or spindle) if two of the $\ell_\alpha \to 0$ and the other approaches infinity; *iii) pancake* if two of the $\ell_\alpha$ approach finite numbers and the other tends to zero as the singularity is approached. In addition to these, we dub as *cylinder* the special case of two of the $\ell_\alpha \to 0$ and the other approaching a constant value.

## 2.2 Silent universes

Having now introduced all the relevant variables we need to describe the dynamics of irrotational dust with vanishing magnetic tidal field, $p = \omega_{ab} = H_{ab} = 0$; they are: $\rho$, $\Theta$, $\sigma_1$, $\sigma_2$, $E_1$ $E_2$. These quantities can be seen as components of a 6–dimensional (6–D) position vector $\vec{X}$ in the phase–space PS6 $= \{\rho, \Theta, \sigma_1, \sigma_2, E_1, E_2\}$. It was indeed shown by Matarrese, Pantano, & Saez (1993) that the dynamics of irrotational dust with zero magnetic tidal field is described by the six first–order *ordinary* differential equations giving the evolution of these quantities along each flow line. After the time evolution of $\Theta$ and the $\sigma_\alpha$'s has been determined by these equations, that of the $\ell_\alpha$'s will be given by integration of Eq.(5). Thus, the history of each fluid element – being given by ordinary differential equations – proceeds with no influence from the environment other than those coded in the initial conditions, as in the case of linear perturbations of a matter dominated FRW universe (e.g. Ellis & Bruni 1989), or in the case of the Zel'dovich approximation in Newtonian theory (Zel'dovich 1970). Since there is no communication between neighborhood fluid flow lines, we may term these as silent universes (Matarrese, Pantano, & Saez 1994a,b).

We may think of the six equations for our variables as a 6–D flow of the form $\dot{\vec{X}} = \vec{V}(\vec{X})$, where $\vec{V}$ at each point $\vec{X}$ is the tangent to the trajectories in the phase–space PS6. The flow $\vec{V} = \vec{V}(\vec{X})$ is a non linear function of $\vec{X}$ only, i.e. $\vec{V}$ does not depend explicitly on time, so that the trajectories do not intersect. Thus, we deal with the following autonomous system:

$$\dot{\rho} = -\Theta\rho , \tag{7}$$

$$\dot{\Theta} = -\frac{1}{3}\Theta^2 - 2\,{\sigma_1}^2 - 2\,\sigma_1\,\sigma_2 - 2\,{\sigma_2}^2 - \frac{1}{2}\rho , \tag{8}$$

$$\dot{\sigma_1} = \frac{2}{3}\sigma_2\,(\sigma_1+\sigma_2) - \frac{1}{3}{\sigma_1}^2 - \frac{2}{3}\,\Theta\,\sigma_1 - E_1 , \tag{9}$$

$$\dot{\sigma_2} = \frac{2}{3}\sigma_1\,(\sigma_1+\sigma_2) - \frac{1}{3}{\sigma_2}^2 - \frac{2}{3}\,\Theta\,\sigma_2 - E_2 , \tag{10}$$

$$\dot{E_1} = E_1\,(\sigma_1-\sigma_2) - E_2\,(\sigma_1+2\,\sigma_2) - \Theta\,E_1 - \frac{1}{2}\rho\,\sigma_1 , \tag{11}$$

$$\dot{E_2} = E_2\,(\sigma_2-\sigma_1) - E_1\,(\sigma_2+2\,\sigma_1) - \Theta\,E_2 - \frac{1}{2}\rho\,\sigma_2 . \tag{12}$$

The first of these equations is the matter conservation, the second is Raychaudhuri equation, while the other two pairs give the time evolution of the shear and the electric tidal field. Using (5) and (6) Eq.(7) gives

$$\rho = \frac{M}{\ell_1\ell_2\ell_3} , \tag{13}$$

as usual, where $M = \rho_* \ell_*^3$, and from now on a subscript $_*$ will label quantities at an arbitrary time $t_*$. As long as one of the $\ell_\alpha$'s vanishes, one has a density singularity. In the Newtonian theory the case of two of the $\ell_\alpha$'s contemporarily going to zero is regarded as exceptional, thus, the general case is that of Zel'dovich pancaking (e.g. Zel'dovich & Novikov 1983, Shandarin & Zel'dovich 1989, Peebles 1993).

However, the relativistic point of view is different, as what really matters are singularities in the curvature. Moreover, for simple homogeneous models the collapse phase is Kasner–like, so that two of the $\ell_\alpha$'s go to zero together, and spindle–like singularities are generic, while the pancake is exceptional (e.g. Zel'dovich & Novikov 1983, Stephani 1990).

For the inhomogeneous Szekeres models it is known that a cigar singularity can also occur (Goode & Wainwright 1982): however, this is rather obvious: since these models have two equal shear eigenvalues, the case for the spindle seems to be as likely as that for the pancake. Then, the question is what is the generic collapsing configuration in the triaxial case: here we will show that the spindle–like singularity is generic in silent universes, since in the last stage of collapse (as well as close to the initial singularity) they tend to have a Kasner–like behaviour.

Although the first work dealing with irrotational fluids with vanishing magnetic Weyl tensor is that of Barnes & Rowlingson (1989), these authors never discussed their cosmological implications. The first cosmological implementation of these models is due to Matarrese, Pantano, & Saez (1984), who also showed the existence of spherically symmetric (Tolman–Bondi) and planar (Zel'dovich) pancake solutions, these solutions arising in a special case in which $\sigma_{ab}$ and $E_{ab}$ are degenerate, i.e. they have two equal eigenvalues. Later, Croudace et al. (1994) discovered an instability of the pancake solution against non–degenerate perturbations, and suggested that such an instability could be ascribed to the disregarding of the magnetic tidal component. Bertschinger & Jain (1994) soon realized that this instability was actually caused by the non–linear tide–shear coupling term in the electric tide evolution equation, which is stabilizing for spindle–like collapse but generally destabilizing for pancakes, except for specific initial data. Bertschinger & Jain (1994), however, noticed that this result was in contradiction with the standard analysis of the Newtonian collapse of isolated ellipsoids (e.g. White & Silk 1979). Matarrese, Pantano, & Saez (1994a) argued that the non–linear tide–shear coupling term is a peculiarity of the relativistic equations, which gives the dominant effect whenever $H_{ab}$ is disregarded: by a second–order perturbative calculation (see also Matarrese, Pantano, & Saez 1994b, Kojima 1994) they showed that, on scales smaller than the horizon size, where the Newtonian approximation should apply, the magnetic tidal tensor cannot be neglected in the general case. Only for perturbations on super–horizon scales the $H_{ab} = 0$ condition would apply, leading to the preferential collapse of fluid elements to spindles, independently of the environmental conditions. The problem has been definitely solved by Kofman & Pogosyan (1994), who obtained the Newtonian limit of the covariant general relativistic equations by a $1/c$ expansion. In particular, they showed that the electric tide evolution equation necessitates a calculation at $1/c^3$ order, in which case a non–zero magnetic tidal field component arises as a post–Newtonian effect, related to non–local gravitational interactions. They also reached the important conclusion that the magnetic Weyl component implies non–local terms, and in the Newtonian limit precisely cancels out the tide–shear coupling term in the tide evolution equation.

Croudace et al. (1994, see also Kasai 1992, 1993, and Salopek, Stewart, & Croudace 1994) pointed out that the planar solution was one of those found by Szekeres (1975a), and examined the dynamics arising in the limit $\Theta \to -\infty$ under the approximation of neglecting linear terms in the equations. They also used a set of variables introduced by Matarrese, Pantano, & Saez (1993), which are defined as non–linear perturbations with respect to a flat FRW background. This point of view is however rather restrictive, and will not be adopted here: it can always be taken at the end of the analysis.

As we will show in the next section, it is convenient to introduce a set of new dimensionless variables in order to study the dynamics of silent universes. However, some information can be already extracted from the system above, before introducing these new variables.

First, we see that the divergence $div\vec{V} = -5\,\Theta$ of $\vec{V}$ depends just on $\Theta$, so that the system is dissipative when the fluid element is expanding ($\Theta > 0$), and exploding otherwise. From the especially simple form Raychaudhuri equation takes for irrotational dust, Eq.(8), we see that $\dot{\Theta} < 0$: thus, once $\Theta < 0$ the collapse of the fluid element is irreversible. The only stationary point[3] of the above system is the origin $O = \{\rho = 0,\, \Theta = 0, \sigma_1 = 0, \sigma_2 = 0, E_1 = 0, E_2 = 0\}$, representing an unstable Minkowski vacuum. Therefore, in the collapse $|\vec{X}| \to \infty$, i.e. the magnitude of all the six variables of the above system grows unbounded.

The system above admits various subsystems describing special subcases. The most obvious one is that of the vacuum, $\rho = 0$. Another one is that of conformally flat models: these are given by $E_1 = E_2 = 0$ ($H_{ab} = 0$ is our fundamental assumption), a condition that can be maintained only in vacuum ($\rho = 0$), or if the fluid is shear–free ($\sigma_1 = \sigma_2 = 0$). Finally, we point out that these conformally flat vacuum models, $\rho = E_1 = E_2 = 0$, correspond to Minkowski spacetime in a disguised form. The most interesting subcase is that of the simultaneous degeneracy of $\sigma_{ab}$ and $E_{ab}$, and will be discussed in the next section.

# 3 Dynamics of silent universes

In this section we will reformulate the dynamical problem for silent universes in terms of new variables that will allow us to simply predict the final fate both of expanding voids and of collapsing configurations, as well as the type of the initial singularity. This formulation will also make evident a sort of time reversal property of the models.

## 3.1 Dimensionless variables

We start by making the following linear transformation in the phase–space PS6:

$$\sigma_\pm = \frac{1}{2}(\sigma_1 \pm \sigma_2)\ , \qquad E_\pm = \frac{1}{2}(E_1 \pm E_2)\ . \tag{14}$$

We then have a new flow $\dot{\vec{Y}} = \vec{\mathcal{V}}(\vec{Y})$ in PS6 $= \{\rho,\, \Theta,\, \sigma_+,\, \sigma_-,\, E_+,\, E_-\}$, given by the 6–D system

$$\dot{\rho} \;=\; -\Theta\,\rho\,, \tag{15}$$

[3] A stationary point of a system of differential equations such as (7)–(12), $\dot{\vec{X}} = \vec{V}(\vec{X})$, is a point $\vec{X}_S$ in phase–space such that $\vec{V}(\vec{X}_S) = 0$. Such a point represents a special solution of the system, and for our purposes here we may say that it can be *asymptotically stable*, simply *stable*, *unstable*, or a *saddle*. An asymptotically stable point will attract generic solutions of the system, i.e. it will be an *attractor* for the trajectories in phase–space, while trajectories around a simply stable point would go around it, without falling on it (e.g. the potential minimum in the mathematical pendulum). A saddle point can be reached if very special initial conditions are chosen (a set of measure zero in phase–space), an unstable point usually represents an asymptotically initial state, i.e. is a *repeller* for the trajectories in phase–space (e.g. Arrowsmith & Place 1982, Arnol'd 1992).

$$\dot{\Theta} = -\frac{1}{3}\Theta^2 - 6\,{\sigma_+}^2 - 2\,{\sigma_-}^2 - \frac{1}{2}\rho \; , \tag{16}$$

$$\dot{\sigma}_+ = {\sigma_+}^2 - \frac{1}{3}{\sigma_-}^2 - \frac{2}{3}\,\Theta\,\sigma_+ - E_+ \; , \tag{17}$$

$$\dot{\sigma}_- = -2\,\sigma_+\,\sigma_- - \frac{2}{3}\,\Theta\,\sigma_- - E_- \; , \tag{18}$$

$$\dot{E}_+ = \sigma_-\,E_- - 3\,E_+\,\sigma_+ - \Theta\,E_+ - \frac{1}{2}\rho\,\sigma_+ \; , \tag{19}$$

$$\dot{E}_- = 3\,\sigma_-\,E_+ + 3\,E_-\,\sigma_+ - \Theta\,E_- - \frac{1}{2}\rho\,\sigma_- \; . \tag{20}$$

Again, $div\vec{\mathcal{V}} = -5\,\Theta$ and the only stationary point is the origin. The interesting subcase is now apparent, and given by $\sigma_- = 0$, which also implies $E_- = 0$: these degenerate ($\sigma_1 = \sigma_2$ and $E_1 = E_2$) models are those of Szekeres (1975a). The above system is obviously symmetric under the simultaneous change of sign of $\sigma_-$ and $E_-$, corresponding to the exchange of axes 1 and 2; it reduces to a 4–D system for $\sigma_- = E_- = 0$: the dynamics of these Szekeres models will be considered in some detail in the next section. Here, we note that, since this dynamics takes place on a 4–D subspace SZ4 of the whole phase–space PS6, trajectories in the latter can "go around" SZ4. Even with the above mentioned symmetry, this means that we cannot restrict the analysis of the dynamics in PS6 to only a part of it unless we also appropriately restrict the initial conditions. Finally, we point out that the degenerate Szekeres models also appear under the less obvious restrictions $\sigma_+ = \pm\frac{1}{3}\sigma_-$ and $E_+ = \pm\frac{1}{3}E_-$ in the system above. These configurations are replicas of the case $\sigma_- = E_- = 0$ which correspond to either $\sigma_1 = \sigma_3$ and $E_1 = E_3$, or $\sigma_2 = \sigma_3$ and $E_2 = E_3$.

In both cases of the two systems (7)–(12) and (15)–(20), the interesting dynamics takes place at infinity in PS6. One should therefore introduce a 6–D Poincaré sphere and a related set of mathematically convenient variables, but this would make the analysis of the dynamics rather abstract. Instead, we prefer to introduce a set of new scalar variables $\Omega$, $\Sigma_+$, $\Sigma_-$, $\varepsilon_+$, $\varepsilon_-$ directly related to physical observables ($\Omega$ for example is just the standard density parameter). These quantities are *dimensionless*, and related to the previous variables by

$$\rho = \frac{1}{3}\Omega\,\Theta^2 \; , \quad \sigma_\pm = \Sigma_\pm\,\Theta \; , \quad E_\pm = \varepsilon_\pm\,\Theta^2 \; ; \tag{21}$$

similar variables have been used in a perturbative context (Goode 1989, Bruni & Piotrkowska 1994). These relations are obviously singular for $\Theta = 0$, but this is not a fact of major concern: it is just a sign of the turn–around $\Theta = 0$, as it is the case for example for $\Omega$ in a closed FRW model. Thus, these new variables will not be very useful to study the turn–around epoch, but they prove to be a good choice to study the collapse and the initial singularities. Moreover, the advantage of using them is in the simplifications occurring in the dynamics, which is now given by a system of the form

$$\dot{\Theta} = -\Theta^2\left(\frac{1}{3} + 6\,{\Sigma_+}^2 + 2\,{\Sigma_-}^2 + \frac{1}{6}\Omega\right) \; , \tag{22}$$

$$\dot{\vec{G}} = \Theta \vec{F}(\vec{G}) , \tag{23}$$

where $\vec{G}$ is a position vector in the 5–D phase–space PS5 = $\{\Omega, \Sigma_+, \Sigma_-, \varepsilon_+, \varepsilon_-\}$. Now the expansion $\Theta$ is factored out in the system above, with the only exception of the Raychaudhuri equation (22), where we got a $\Theta^2$ factor. The origin is again a stationary point, but in addition to it we now have a whole 5–D stationary hyperplane $\Theta = 0$. However, we have to cut out this plane from our analysis, as our new variables $\Omega$, $\Sigma_+$, $\Sigma_-$, $\varepsilon_+$, $\varepsilon_-$ diverge on it. Instead, we will now split the analysis of the dynamics of our models in two parts, one for $\Theta > 0$ and one for $\Theta < 0$. In order to achieve this splitting, we introduce a new "time" variable

$$\tau = \pm \int \Theta dt = \pm 3 \ln \ell , \tag{24}$$

where we use the plus sign in the above definition when $\Theta > 0$, and the minus sign when $\Theta < 0$. This choice ensures that $d\tau/dt > 0$ whatever is the sign of $\Theta$. Moreover, $\tau$ is a convenient time variable in that $\tau \to \infty$ in getting close to the collapse or the initial singularity, when $\Theta \to -\infty$, as $\ell \to 0$. Moreover, $\tau \to \infty$ also signals the complete evacuation of an ever–expanding configuration: in such a case $\Theta \to 0$, but $\ell \to \infty$. Denoting by a prime the derivative with respect to $\tau$, the evolution equations for our variables for $\Theta < 0$ read

$$\Theta' = \Theta \left( \frac{1}{3} + 6\,{\Sigma_+}^2 + 2\,{\Sigma_-}^2 + \frac{1}{6}\Omega \right) , \tag{25}$$

$$\Omega' = -\frac{1}{3}\Omega \left( 36\,{\Sigma_+}^2 - 1 + 12\,{\Sigma_-}^2 + \Omega \right) , \tag{26}$$

$$\Sigma_+' = \Sigma_+ \left[ \frac{1}{3} - \Sigma_+ (1 + 6\,\Sigma_+) - 2\,{\Sigma_-}^2 - \frac{1}{6}\,\Omega \right] + \frac{1}{3}{\Sigma_-}^2 + \varepsilon_+ , \tag{27}$$

$$\Sigma_-' = \Sigma_- \left[ \frac{1}{3} - 2\,\Sigma_+ \,(3\,\Sigma_+ - 1) - 2\,{\Sigma_-}^2 - \frac{1}{6}\,\Omega \right] + \varepsilon_- , \tag{28}$$

$$\varepsilon_+' = \varepsilon_+ \left[ \frac{1}{3} - 3\,\Sigma_+ \,(4\,\Sigma_+ - 1) - 4\,{\Sigma_-}^2 - \frac{1}{3}\,\Omega \right] - \Sigma_- \,\varepsilon_- + \frac{1}{6}\Sigma_+ \,\Omega , \tag{29}$$

$$\varepsilon_-' = \varepsilon_- \left[ \frac{1}{3} - 3\,\Sigma_+ \,(4\,\Sigma_+ + 1) - 4\,{\Sigma_-}^2 - \frac{1}{3}\,\Omega \right] - 3\,\Sigma_- \,\varepsilon_+ + \frac{1}{6}\Sigma_- \,\Omega \tag{30}$$

We see that, by the introduction of the dimensionless variables $\Omega$, $\Sigma_\pm$ and $\varepsilon_\pm$ and the time $\tau$, we have achieved many important simplifications: *i)* the equations for $\Omega$, $\Sigma_\pm$ and $\varepsilon_\pm$ *do not depend on the expansion* $\Theta$; *ii)* this means that we have achieved a *dimensional reduction*, since we can now analyse the dynamics of the 5–D flow $\vec{G}\,' = \pm \vec{F}(\vec{G})$ (+ for expansion, − for collapse) given by the subsystem (26)–(30); *iii)* the change of sign of the expansion $\Theta$ corresponds now to a "time reversal" $\tau \to -\tau$, under which the only change in the equations above is the sign change of the right hand side; *iv)* this means that the dynamics is completely specular under this time reversal, in the sense that the trajectories in PS5 are exactly the same for $\Theta < 0$ and $\Theta > 0$, as only the direction of the tangent to the trajectories is changed for $\tau \to -\tau$, i.e. $\Theta \to -\Theta \Leftrightarrow \vec{G}\,' \to -\vec{G}\,'$. As we will see, stationary points in PS5 will now appear for *finite values* of $\Omega$, $\Sigma_\pm$, and $\varepsilon_\pm$: because of *i)* these stationary points will be the same for $\Theta > 0$ and $\Theta < 0$, and because of *i)* and *iv)* the Jacobian $J = \partial\vec{G}\,'/\partial\vec{G}$ changes sign with $\Theta$: in particular, the eigenvalues of $J$ change sign, so that stable points (attractors) become unstable (repellers), and completely unstable points (repellers) become stable (attractors). Saddle points of course remain saddles. The Raychaudhuri equation (25) is now merely a point dependent clock in PS5 (i.e. the time $\tau$ (24) depends on the point $\vec{G}$ in PS5), and it can be integrated at stationary points. Because of the $\Theta$ factor in the right hand side of (25) and considering the definition of $\tau$, Eq.(24), $\Theta'$ *will be the same under the time reversal* $\tau \to -\tau \Leftrightarrow \Theta \to -\Theta$ at every point $\vec{G}$ of PS5. This is also true for $\dot{\Theta}$, as it is obvious from (22).

In synthesis, the phase–space during collapse is a mirror image of the phase–space during expansion, with the mirror reversing the arrow of the time $\tau$.

## 3.2 Stationary points

As we already said, stationary points will now appear for finite (constant) values of $\Omega$, $\Sigma_\pm$, $\varepsilon_\pm$; thus, denoting by a subscript $_S$ these values, the Raychaudhuri equation (25) at stationary points reduces to

$$\Theta' = \pm\beta\Theta\ , \qquad \beta = \frac{1}{3} + 2(3\Sigma_{+S}^{\,2} + \Sigma_{-S}^{\,2}) + \frac{1}{6}\Omega_S\ , \tag{31}$$

where here and in what follows the top sign holds for the collapse phase $\Theta < 0$ and the bottom sign for the expansion phase $\Theta > 0$. This gives

$$\Theta = \Theta_*\, e^{\pm\beta(\tau-\tau_*)}\ , \tag{32}$$

and

$$e^{(\tau-\tau_*)} = [1 + \beta\Theta_*(t - t_*)]^{\mp\frac{1}{\beta}}\ ; \tag{33}$$

this relation between $\tau$ and $t$ clarifies that $\tau \to \infty$ when a singularity is approached. For $\Theta$, $\ell$ and the $\ell_\alpha$'s we have

$$\begin{aligned} \Theta &= \frac{\Theta_*}{1 + \beta\Theta_*(t - t_*)}\ , & (34)\\ \ell &= \ell_*[1 + \beta\Theta_*(t - t_*)]^{\frac{1}{3\beta}}\ , & (35)\\ \ell_\alpha &= \ell_{\alpha *}[1 + \beta\Theta_*(t - t_*)]^{p_\alpha}\ , & (36) \end{aligned}$$

where $p_\alpha = \frac{1}{\beta}(\frac{1}{3} + \Sigma_\alpha)$ and $\Sigma_1 = \Sigma_+ + \Sigma_-$, $\Sigma_2 = \Sigma_+ - \Sigma_-$, $\Sigma_3 = -2\Sigma_+$. From these relations we see that the expanding or contracting behaviour of each model given by a stationary point is monotonic and fixed by the sign of $\Theta_*$. In dependence of this sign there is a singularity, either in the past (initial singularity for $\Theta_* > 0$) at $t - t_* = -(\beta\Theta_*)^{-1}$, or in the future ($\Theta_* < 0$) at $t - t_* = (\beta|\Theta_*|)^{-1}$.

Before proceeding with the analysis of the more complicated 5–D system (26)–(30), we consider the degenerate case ($\Sigma_- = \varepsilon_- = 0$) of Szekeres models. These will now be described by a 3–D flow $\vec{g}\,' = \pm\vec{f}(\vec{g})$ (again, + for expansion, − for collapse) in the phase–space SZ3 = $\{\Omega,\ \Sigma_+,\ \varepsilon_+\}$. This flow can be visualized, and this will be instructive also for the most general case.

# 4 Dynamics of Szekeres models

That the irrotational dust models with vanishing magnetic tidal field and with degenerate (two equal eigenvalues) shear and electric tidal field must be the solutions of Szekeres (1975a) was shown by Barnes & Rowlingson (1989).

Szekeres models generalize Kantowski–Sachs, FRW, and Tolman–Bondi solutions (Szekeres 1975a) and were studied in great detail by many authors because of their interest as inhomogeneous solutions of Einstein equations with no symmetries (Kramer et al. 1980; MacCallum 1993). In particular, Szekeres (1975b) considered the collapse of a subclass with finite mass; Bonnor (1976) showed that these spacetimes can be matched to a spherically symmetric Schwarzschild vacuum solution; Bonnor & Tomimura (1976) studied the time evolution of the most cosmologically interesting subclass; Lawitzky (1980) gave their Newtonian analogs; Barrow & Silk (1981) considered them in their study of the growth of anisotropic structures in the universe; Goode & Wainwright (1982) emphasized how these models can be characterized by the growing and decaying modes of dust–filled FRW models, and studied the character of their singularities in detail (Goode & Wainwright 1992, and references therein).

Although Szekeres models are usually divided in two classes, their evolution along flow lines is the same in both of them, as noticed by Goode & Wainwright (1982). Since we are precisely interested in this aspect of the models, we will not need to distinguish between the two classes.

## 4.1 An overview of phase–space

During the collapsing phase $\Theta < 0$ the dynamics of Szekeres models is given by setting $\Sigma_- = \varepsilon_- = 0$ in (28), (30). Then, we obtain an autonomous 3–D subsystem in SZ3, $\vec{g}\,' = -\vec{f}(\vec{g})$, i.e.

$$\Omega' = -\frac{1}{3}\Omega\left(36\,\Sigma_+{}^2 - 1 + \Omega\right)\,, \tag{37}$$

$$\Sigma_+' = \Sigma_+\left[\frac{1}{3} - \Sigma_+\left(1 + 6\,\Sigma_+\right) - \frac{1}{6}\,\Omega\right] + \varepsilon_+\,, \tag{38}$$

$$\varepsilon_+' = \varepsilon_+\left[\frac{1}{3} - 3\,\Sigma_+\,\left(4\,\Sigma_+ - 1\right) - \frac{1}{3}\,\Omega\right] + \frac{1}{6}\Sigma_+\,\Omega\,, \tag{39}$$

while, at each point $\vec{g}$ of SZ3, the Raychaudhuri equation gives $\tau = \tau(\vec{g})$:

$$\Theta' = \Theta\left(\frac{1}{3} + 6\,\Sigma_+{}^2 + \frac{1}{6}\Omega\right)\,. \tag{40}$$

The divergence $div\vec{f}$ of the flow $\vec{f}$ in SZ3 is such that

$$div\vec{f} < 0 \qquad \text{for} \qquad \Omega > \frac{6}{7}\left(1 + \Sigma_+ - 42\,{\Sigma_+}^2\right) . \tag{41}$$

Thus, for $\Theta < 0$ the flow $\vec{f}$ is converging everywhere in SZ3, except below the parabola given in (41). Remarkably, we see from (41) that $div\vec{f}$ does not depend on $\varepsilon_+$. From (37) we see that the trajectories in SZ3 cannot cross the $\Omega = 0$ plane (because of the $\Omega$ factor in the right hand side), which corresponds to the fact that only the $\Omega \geq 0$ section of SZ3 is physically meaningful. Also, from (37) we see that $\Omega' < 0$ for $\Omega > 1 - 36\,{\Sigma_+}^2$, i.e. everywhere (41) is also satisfied. Then, from these three facts we may conclude that for $\Theta < 0$ *generic trajectories in* SZ3 *converge toward the* $\Omega = 0$ *plane*, i.e. we may expect to find stationary points, in particular attracting ones, on this plane.

It is immediate to find these points $\vec{g}_S$ with a computer algebra system: they are listed in Table 1, giving the coordinate of each point in the phase–space SZ3. With a little more work one can easily compute the Jacobian $J(\vec{g}_S) = -[\partial\vec{f}/\partial\vec{g}](\vec{g}_S)$ at each of these points, and then find its eigenvalues $\lambda_i$ ($i = 1, 2, 3$). These eigenvalues are listed in Table 2, with the values they have during collapse, $\Theta < 0$. Points with all $\lambda_i < 0$ are *asymptotically stable*, i.e. the generic trajectory in SZ3 is attracted by one of these points. Conversely, points with all $\lambda_i > 0$ are *unstable*. Finally, points with at least two of the $\lambda_i$'s having different sign are saddles, and are therefore unstable.

As previously explained in Section 3, the right hand side of Eqs.(37)–(39) changes sign under the time reversal $\tau \to -\tau$, i.e. in considering the expanding phase, $\Theta > 0$. Then, also the Jacobian $J(\vec{g})$ changes sign, together with its trace $div\vec{f}$ and eigenvalues $\lambda_i$. Therefore, the inequality (41) is reversed in the expanding phase, and the generic trajectories escape from the $\Omega = 0$ plane. Since the $\lambda_i$ have different sign during expansion and collapse, the nature of the stationary points also changes. In the last two columns of Table 2 we have therefore given the type of each point during collapse and during expansion. To further illustrate the behaviour of trajectories in phase–space SZ3, we have plotted in Figure 1 the vector flow $\vec{f}$ above points D II and D V, during the collapse phase, each of these points being at the center of the bottom of the plotted boxes. From these figures it is clearly seen how during collapse point D II is repelling, although the flow $-\vec{f}$ above it is directed downward for $\Omega \gtrsim 1$; the flow $-\vec{f}$ above point D V is directed toward it, and D V is attracting.

Figure 2 (a) represents the Poincaré disk for the whole $\Omega = 0$ plane [4] during collapse, $\Theta < 0$, i.e. the bottom of the boxes in Figure 1, and Figure 2 (b) represents the central part of the same plane including all the five stationary points D II–D VI, for $\Theta > 0$. A comparison between the top views in Figure 1 and Figure 2 reveals how, above each point and for $\Omega$ small enough, the trajectories are basically the same as for $\Omega = 0$.

## 4.2 Interpretation of the stationary points

We now give an interpretation of the stationary points listed in Table 1; again, we remark that they are the same for expansion ($\Theta > 0$) and collapse ($\Theta < 0$). As mentioned in Section 3.2, in general the stationary points represent models that either expand forever from an initial singularity (when we chose $\Theta > 0$), or that collapse from infinity to a future singularity (for $\Theta < 0$, i.e. the latter are true time reversal of the former ones). The values of $\beta$ and of the exponents $p_\alpha$'s in (36) are given in Table 3. Then, point D I represents a flat FRW model, having zero shear and tidal field. Point D II clearly represents an empty conformally flat shear–free void, locally equivalent to an empty open FRW model (a Milne universe). Point D III is a vacuum solution we have not been able to identify in the literature (see below). Using its $\beta$ and $p_\alpha$ values, it is easy to check that Point D IV represents the degenerate Kasner model with two non–expanding directions, and a pancake singularity. Similarly, point D V represents another degenerate Kasner model, with two equally expanding directions and a contracting one, with a cigar singularity. Point D VI is the limit of a subclass of Szekeres models (see below, and Section 5). Point D VII is clearly unphysical, having $\Omega < 0$. Notice that the points D II, D III and D IV are locally equivalent to Minkowski, (M) in the Tables. In general, configurations with $\Sigma_+ > 0$ ($< 0$) represent prolate (oblate) spheroidal fluid elements during collapse (the inequalities must be reversed during expansion).

[4] This disk is a conformal map of the $\{\Sigma_+ , \varepsilon_+\}$ plane, so that its boundary represents infinity in this plane.

Now, more interesting than the stationary points themselves is the behaviour of trajectories in their neighbourhood in phase–space. For Szekeres models, this is given by the eigenvalues of the Jacobian $J(\vec{g}_S) = -[\partial \vec{f} / \partial \vec{g}](\vec{g}_S)$ at each of the stationary points $\vec{g}_S$ of the system (37)–(39): these eigenvalues are listed in Table 2, for the collapsing phase $\Theta < 0$. Points D I, D III, D VI are saddles, i.e. are unstable in any case, except for special initial conditions. For example, during expansion the FRW point, D I, can be approached only by initial conditions corresponding to the pure decaying mode of linear perturbation theory (see Section 5.2). Point D II is a repeller (an unstable star node) during collapse, and an attractor (a stable star node) during expansion: *it represents the final fate of ever–expanding voids*, locally equivalent to an empty open FRW (Milne) universe. Point D III is a saddle in the $\{\Sigma_+, \varepsilon_+\}$ plane, and has a vanishing eigenvalue for the direction out of this plane: it is a vacuum solution we have not been able to identify in the literature, but being a saddle and representing Minkowski, it seems of no interest. The Kasner points D IV and D V are repellers (unstable nodes) during expansion, and attractors (stable nodes) during collapse. Therefore, point D IV represents the final fate of locally axisymmetric pancakes, and point D V that of filaments. Point D VI is a saddle, but during expansion it has only one positive eigenvalue: thus, there is a subset of trajectories (of measure zero) that fall on it; this point indeed corresponds to the limit $t \to \infty$ of Szekeres models P I by Bonnor & Tomimura (1976), i.e. there is a subset of special Szekeres models that expand and asymptotically tend to D VI. The unphysical point D VII is unstable.

Let us look again to Figure 2: the two plots show that the trajectories are the same during collapse and expansion, but the directions are reversed: the two attractors D IV and D V in the upper figure are unstable in the bottom figure, for $\Theta > 0$, while the origin is now stable; it represents the final fate of voids. Note that the $\varepsilon_+ = 0$ axis is a separatrix between the $\varepsilon_+ > 0$ and the $\varepsilon_+ < 0$ worlds; since $H_{ab} = 0$ by assumption, and $\rho = 0$ on the plane in figure, the $\varepsilon_+ = 0$ line represents Minkowski spacetime: in particular the three points on the line from left to right are D IV, D II, and D III. The two points above the $\varepsilon_+ = 0$ axis are D VI (left) and D V (right).

Finally, Figure 3 shows the Poincaré disk for the the subcase $\varepsilon_+ = \Omega = 0$, i.e. for Minkowski

models mentioned above, appearing on the $\varepsilon_- = 0$ line in Figure 2. Since only $\Sigma_+$ varies for these models, one can resort to the variables $\sigma_+$ and $\Theta$, thus showing the transition from expansion to collapse. Although the trajectories in the resulting $\{\sigma_+, \Theta\}$ plane are just Minkowski spacetime in disguised form, they give an idea of the general behaviour.

The central stationary point is Minkowski in its standard form (static and shear free). In these conformal maps straight lines stay straight: the vertical $\sigma_+ = 0$ axis is Milne universe, point D II (expanding in the upper half, contracting in the lower half); the two lines $\Theta = 6\sigma_+$ and $\Theta = -3\sigma_+$ correspond to points D III and DIV respectively. Trajectories enclosed between these two lines in the upper half (lower half) of the figure schematically represent the behaviour of ever expanding (contracting) voids. Trajectories outside the two lines represent configurations that first expand, then recollapse. At infinity all the trajectories asymptotically approach the two lines.

### 4.3 Szekeres trajectories

In Section 5 we will give the asymptotic behaviour around points D I–D VI of the generic trajectories representing triaxial configurations.

In order to show how the solution of the general system of Eqs.(7)–(12) could be possibly afforded, we briefly report here on the solution of the restricted system for the degenerate case $\sigma_- = 0$ and $E_- = 0$, i.e. for Szekeres models. We will basically follow the analysis by Goode & Wainwright (1982), but we will only focus on the solution of the time evolution equations, without paying attention to the spatial dependence of the various quantities, i.e. without dealing with the spatial constraint equations. Note that our notation here will be slightly different to that adopted by Goode & Wainwright (1982).

Up to a scale change which is constant along the flow lines, we can write

$$\ell_1(\vec{x}, t) = \ell_2(\vec{x}, t) \equiv S(\vec{x}, t) \ , \tag{42}$$

and

$$\ell_3(\vec{x}, t) \equiv S(\vec{x}, t) Z(\vec{x}, t) \ , \tag{43}$$

which implies $\rho = M/S^3 Z$, with $M$ a generally space–dependent integration constant, and

$$\Theta = 3\frac{\dot{S}}{S} + \frac{\dot{Z}}{Z} \ , \qquad \sigma_+ = -\frac{1}{3}\frac{\dot{Z}}{Z} \ . \tag{44}$$

Replacing these expressions into the tide evolution equation, we obtain

$$E_+ = -\frac{M}{6S^3 Z} + \frac{N}{S^3} \ , \tag{45}$$

where $N$ is another (generally space–dependent) integration constant. The evolution equations for the expansion scalar and the shear are simplified by introducing the function $F \equiv Z - \frac{M}{6N}$, which leads to the equation

$$\ddot{F} + 2\frac{\dot{S}}{S}\dot{F} = \frac{3N}{S^3} F \ , \tag{46}$$

and

$$\ddot{S} = -\frac{N}{S^2} \ . \tag{47}$$

The latter equation can be integrated once to give

$$\left(\frac{\dot{S}}{S}\right)^2 = \frac{2N}{S^3} - \frac{k}{S^2} \ , \tag{48}$$

where $k$ is an integration constant; by properly rescaling $S$ and $N$ one can renormalize $k$ in such a way that it can only take the values $-1, 0, +1$.

It is then clear that the function $S$ is equivalent to the scale factor of matter dominated FRW models, in the sense that, for each fluid element, $S$ satisfies the Friedmann equation for an open, flat or closed FRW model, depending on whether $k$ is respectively $-1$, $0$ or $+1$. The function $F$, on the other hand, has to satisfy Eq.(46), which is the equation for linear perturbations around FRW; one has

$$F = -\beta^{(+)} f^{(+)} - \beta^{(-)} f^{(-)} \ , \tag{49}$$

where $f^{(+)}$ and $f^{(-)}$ represent the growing and decaying modes of linear perturbations around FRW and $\beta^{(+)}$, $\beta^{(-)}$ set the corresponding (generally space–dependent) amplitudes.

The solutions of Eq.(48) and (46) can be found e.g. in Peebles (1980), we nevertheless report their form here for completeness. As far as the scale factor is concerned one has

$$S = N h'(\eta) \ , \qquad t - t_0 = N h(\eta) \ , \tag{50}$$

where

$$h(\eta) = \begin{cases} \sinh\eta - \eta \ , & k = -1 \ , \\ \frac{1}{6}\eta^3 \ , & k = 0 \ , \\ \eta - \sin\eta \ , & k = +1 \ ; \end{cases} \tag{51}$$

here a prime denotes differentiation with respect to the *development angle* $\eta$ and $t_0$ is a (generally space–dependent) constant. The growing and decaying modes read

$$f^{(+)} = \begin{cases} \frac{6N}{S}[1 - (\eta/2)\coth(\eta/2)] + 1 \ , & k = -1 \ , \\ \frac{1}{10}\eta^2, & k = 0 \ , \\ \frac{6N}{S}[1 - (\eta/2)\cot(\eta/2)] - 1 \ , & k = +1 \ , \end{cases} \tag{52}$$

and

$$f^{(-)} = \begin{cases} \frac{6N}{S}\coth(\eta/2) \ , & k = -1 \ , \\ 24\eta^3 \ , & k = 0 \ , \\ \frac{6N}{S}\cot(\eta/2) \ , & k = +1 \ . \end{cases} \tag{53}$$

Goode & Wainwright (1982) were able to show that, for $k = -1$ or $0$, the final singularity in Szekeres models can only be pancake–like and corresponds to $Z \to 0$; for $k = +1$ the final singularity is point–like when $\beta^{(+)} = \beta^{(-)}/\pi$ (in correspondence to $S \to 0$), cigar when $\beta^{(+)} < \beta^{(-)}/\pi$ (still for $S \to 0$) and pancake when $\beta^{(+)} > \beta^{(-)}/\pi$ (for $Z \to 0$).

Finally, let us notice that the planar pancake solution obtained by Matarrese, Pantano, & Saez (1993) is recovered in the above formalism, by taking $k = 0 = f^{(-)} = 0$. This particular trajectory undergoes pancake–like collapse to a shell–crossing singularity, asymptotically approaching our attractor D IV. Indeed, one may argue that the only pancake–like collapse to D IV (with its replicas) always leads to shell–crossing, contrary to the most generic spindle one.

# 5 Triaxial dynamics

In this section we study the stationary points of the triaxial general system (26)–(30), with special emphasis on the attracting set, which we will argue to be given by the whole Kasner family of vacuum solutions of general relativity. Then, we will give the asymptotic behaviour of trajectories in the neighborhood of each of the points D I–D VI, for an arbitrary initial condition in this neighborhood in the 5–D phase–space PS5.

## 5.1 Kasner attractor set for triaxial collapse

The triaxial dynamics is similar to that of Szekeres models; again, during collapse trajectories are directed toward the $\Omega = 0$ plane, as (26) implies $\Omega' < 0$, and

$$div\,\vec{G} < 0 \qquad \text{for} \qquad \Omega > 1 - 12\,\Sigma^2\,, \tag{54}$$

where $\Sigma^2 = 3\,{\Sigma_+}^2 + {\Sigma_-}^2$ is the dimensionless shear magnitude. There is however a very important difference: in addition to the points D I–D VII (which obviously are stationary points also for the general case), and their replicas, there is now a whole family of stationary points that form a closed curve in PS5, and which are attractors during collapse.

The new points are listed in Table 4. Again, these points are the same for expansion and collapse. For each listed point T there is a specular one $\overline{\mathrm{T}}$ given by the map $\Sigma_- \to -\Sigma_-$, $\varepsilon_- \to -\varepsilon_-$. More in general there are six distinct replicas (three in T and three in $\overline{\mathrm{T}}$) for each truly triaxial physical configuration, corresponding to the 3! permutations of the principal axes of an ellipsoidal fluid element; each degenerate spheroidal configuration, on the other hand, shows up in 3!/2!=3 replicas.

The eigenvalues $\lambda_i$ of the Jacobian $J = -\vec{F}(\vec{G}_S)$ at each stationary point $\vec{G}_S$ of the system (26)–(30) are given in Table 5. The value of $\beta$ in (36), and the exponents $p_\alpha$, are given in Table 3. Points T I and T II are replicas of D III and D VI respectively. Points of the family T III are parametrized by $-1/3 \le \Sigma_+ \le 1/3$, where the functions in Table 4 read:

$$\Sigma_-(\Sigma_+) \;=\; \frac{1}{\sqrt{3}}\sqrt{1 - 9{\Sigma_+}^2}\,, \tag{55}$$

$$\varepsilon_+(\Sigma_+) \;=\; \frac{1}{3}\Sigma_+(6\Sigma_+ + 1) - \frac{1}{9}\,, \tag{56}$$

$$\varepsilon_-(\Sigma_+) \;=\; -\frac{\sqrt{3}}{9}(6\Sigma_+ - 1)\sqrt{1-9\Sigma_+{}^2}\;; \tag{57}$$

therefore, the family T III is a curve in the 5–D phase–space PS5 (actually, it lies on the hyperplane $\Omega = 0$ in PS5) parametrized by $\Sigma_+$. Let us call $\mathcal{A} = \mathcal{A}(\Sigma_+)$ the curve parametrized by $\Sigma_+$, and including both the T III and the $\overline{\text{T III}}$ family. This curve is projected in the $\{\Sigma_+, \Sigma_-\}$ plane on the ellipse

$$3\,\Sigma^2 = 9\,\Sigma_+{}^2 + 3\,\Sigma_-{}^2 = 1 \qquad \Leftrightarrow \qquad \sigma^2 = \frac{1}{3}\Theta^2\;. \tag{58}$$

This ellipse can obviously be represented in parametric form, using the dimensionless shear magnitude $\Sigma$ and the angle $\alpha$ in the plane $\{\Sigma_+\,, \frac{1}{\sqrt{3}}\,\Sigma_-\}$ (see Appendix A). Similarly, the curve $\mathcal{A}$ in PS5 is projected in the $\{\varepsilon_+\,, \varepsilon_-\}$ plane through (56)–(57), and it may be represented in parametric form using the dimensionless tidal field magnitude $\varepsilon^2 = 3\,\varepsilon_+{}^2 + \varepsilon_-{}^2$ and the angle $\beta$ in the plane $\{\varepsilon_+, \frac{1}{\sqrt{3}}\,\varepsilon_-\}$. Since this curve is parametrized by $\Sigma_+$ only, $\beta$ is a function of $\alpha$; using $\beta = -\frac{1}{2}\alpha$ we get for these curves (the subscript $_{\mathcal{A}}$ labels the values on the curve $\mathcal{A}$):

$$\begin{aligned} \Sigma_+ &= \tfrac{1}{\sqrt{3}}\,\Sigma_{\mathcal{A}}\,\cos(\alpha)\;, \\ \Sigma_- &= \Sigma_{\mathcal{A}}\,\sin(\alpha)\;, \end{aligned} \qquad \text{with} \quad \Sigma_{\mathcal{A}} = \frac{1}{\sqrt{3}}\;, \tag{59}$$

$$\begin{aligned} \varepsilon_+ &= \tfrac{1}{\sqrt{3}}\,\varepsilon_{\mathcal{A}}(\beta)\,\cos(\beta)\;, \\ \varepsilon_- &= \varepsilon_{\mathcal{A}}(\beta)\,\sin(\beta)\;, \end{aligned} \qquad \text{with} \quad \varepsilon_{\mathcal{A}}(\beta) = \frac{2}{3\sqrt{3}}\,\cos(3\beta)\;. \tag{60}$$

These curves are shown in Figure 4, where that in the $\{\varepsilon_+\,, \varepsilon_-\}$ plane is a three–lobe curve with period $\pi$ (the function $\varepsilon_{\mathcal{A}}(\beta)$ is itself a symmetric three–leaf curve with period $\frac{2}{3}\pi$). The fact that the latter intersect itself is a projection effect: the curve $\mathcal{A}$ on $\Omega = 0$ in PS5 is closed with no self–intersections ($\mathcal{A}$ can be visualized in the 3–space $\{\,\varepsilon_+\,, \varepsilon_-\,, \Sigma_+\,\}$ using $\Sigma_+$ as parameter along the curve). The curve $\mathcal{A}$ include all the six possible replicas of each physical configuration in the family T III, $\overline{\text{T III}}$, with degenerate (Szekeres) configurations appearing three times (dots in Figure 4, at the intersections with the lines $\Sigma_- = 0$, $\varepsilon_- = 0$, $\Sigma_- = \pm 3\,\Sigma_+$ and $\varepsilon_- = \pm 3\,\varepsilon_+$ (dashed lines in Figure 4. On the curve $\mathcal{A}$, points T IV, $\overline{\text{T IV}}$ are replicas of D VII: all unphysical stationary points are degenerate. Point T III with $\Sigma_+ = -1/3$ is just D IV, and points T III and $\overline{\text{T III}}$ with $\Sigma_+ = 1/6$ are its replicas: these are the only solutions of the family which end up in a pancake singularity. Point T III with $\Sigma_+ = 1/3$ is just D V, and T III, $\overline{\text{T III}}$, with $\Sigma_+ = -1/6$, are its replicas; these solutions end up in a cigar singularity.

During collapse (expansion) points in T III represent flattened (elongated) ellipsoids for $-1/3 \le \Sigma_+ \le -1/(2\sqrt{3})$ and for $0 \le \Sigma_+ \le 1/(2\sqrt{3})$, and elongated (flattened) ellipsoids for $-1/(2\sqrt{3}) \le \Sigma_+ \le 0$ and for $1/(2\sqrt{3}) \le \Sigma_+ \le 1/3$.

Using in Eqs.(4) and (36) the values of $\beta$ and $p_\alpha$ given in Tab. 3 it is easy to check that the whole family T III, $\overline{\text{T III}}$ locally corresponds to the whole set of Kasner vacuum solutions of general relativity. As it is well known (Zel'dovich & Novikov 1983, Stephani 1990), there is a single pancake singularity in this set (given by D IV and its replicas), while the singularity for the generic Kasner model is spindle–like.

In Table 5 we give the eigenvalues of the Jacobian $J(\vec{G}_S)$ at each of the stationary points $\vec{G}_S$ of the general system (26)–(30), for $\Theta < 0$ (again, the $\lambda$'s change sign for $\Theta > 0$). We include here also the points D I–D VII of the degenerate case listed in Table 1, in order to show their stability properties in the full 5–D phase–space PS5 (cf. Table 2). In particular, the flat FRW point D I is again a saddle, and therefore unstable both during expansion and during collapse; the Milne point D II is again the attractor for expanding voids. Points T I, T II and T IV are equivalent to D III, D VI and D VII respectively, as is made clear by inspection of their $\lambda$'s. Also, points D IV and D V are included in the family T III, and their $\lambda$'s are obtained by those of T III setting $\Sigma_+ = -1/3$, $\Sigma_+ = 1/6$ and $\Sigma_+ = -1/6$, $\Sigma_+ = 1/3$ respectively. Note that point D II is asymptotically stable in the expanding phase (all $\lambda$'s are negative in this case).

All saddle points are degenerate, and have at least two eigenvalues of different sign; the negative ones correspond to eigenvectors in the direction of special trajectories along which that point is approached. For instance, in the collapse phase the closed FRW model and a special Szekeres model end up in point D I.

Points of the family T III have $\lambda_2 = 0$, and $\lambda_i < 0$ $(i \neq 2)$ during collapse. Thus, the outcome of linearization stability analysis is that in the collapsing phase the curve $\mathcal{A}$ is an attractor. Each point on $\mathcal{A}$ given by a value of $\Sigma_+$ is asymptotically stable against a perturbation in a 4–D subspace in PS5, and it is simply stable in the direction of the nearby point given by $\Sigma_+ + \delta\Sigma_+$. In other words, the vanishing of one of the eigenvalue indicates an invariance in the direction tangent to the attractor curve in phase–space, so that for each point of the curve $\mathcal{A}$ there should be a 4–D subspace from which this point is reached. This is in fact the outcome of any numerical test we have done. Then, this 4–D subspace would be made by all those trajectories that end up on that point. Therefore, we conclude that each point on the curve $\mathcal{A}$, locally representing the Kasner set of vacuum homogeneous solutions of general relativity, is an attractor for a set of generically triaxial configurations, which is precisely the set of trajectories forming the 4–D subspace passing through the point. Thus, the generic triaxial configuration that passes through turn–around tends to one point on the curve $\mathcal{A}$, and collapses to a Kasner–like triaxial spindle singularity. Caused by our choice of time variable, given by (33), the approach to the singularity is asymptotic in PS5, as expressed by $\tau \to \infty$, but occurs in a finite proper time $t$. This approach will be further clarified by the asymptotic analysis of next section. The fact that close to the singularity matter is unimportant, as testified by $\Omega \to 0$, should not surprise. Close to the singularity density and expansion are unconnected, so that the density is inhomogeneous, with $\rho \to \infty$ but $\Omega \to 0$, while the metric can even be homogeneous, as shown for example by Bonnor & Tomimura (1976) for a special Szekeres model.

We have to point out that the discussion above, although proving the attracting nature of each single point of the curve $\mathcal{A}$, is in the following sense incomplete. Indeed, one should be able to demonstrate that there is a conserved quantity $Q$ (i.e. $Q' = 0$), parametrized in the same way as the attractor curve $\mathcal{A}$ (e.g. by $\Sigma_+$), that foliates the phase–space PS5. Then, for each point $\vec{G}_{\mathcal{A}}$ on $\mathcal{A}$ there would be a surface passing through it (the 4–D manifold above), and the trajectories lying on this surface would asymptotically fall on $\vec{G}_{\mathcal{A}}$ (the negative eigenvalues at $\vec{G}_{\mathcal{A}}$ corresponding to motion in the surface); the eigenvector of $J(\vec{G}_{\mathcal{A}})$ corresponding to the vanishing eigenvalue, being tangent to the curve, would be transverse (not necessarily orthogonal) to these surfaces, and would map the motion from one surface to the next one. We stress again that, although we have not been able to show the existence of such a conserved quantity $Q$, numerical tests indicate that it should exist.

## 5.2 Asymptotic behaviour around stationary points

From Eq.(36) one can see that the behaviour of the $\ell_\alpha$'s is similar to that of Kasner models for point D III, while they all expand or contract for D VI (they are trivially the same for D I and D II), and all other points are in the T III family, i.e. they are Kasner models.

We can now give the asymptotic behaviour of the general solutions of the system (26)–(30) in the neighborhood of each of these points. This is achieved by using the Jacobian of the system at each stationary point, solving the linear problem $\vec{G}\,' = J(\vec{G}_S)\vec{G}$ with an arbitrary initial condition $\vec{G}_*$, where however $\vec{G}_*$ is supposed to be close to $\vec{G}_S$. In doing so, we solve a direct problem, i.e. we answer the question of where a trajectory starting in the neighborhood of $\vec{G}_S$ is going to end up to. We leave on a side the inverse problem, i.e. that of determining the set of initial data in the neighborhood of $\vec{G}_S$ from which trajectories would fall on $\vec{G}_S$ itself.

For all points except D I, D II and D VI we give here the behaviour for collapse, $\Theta < 0$, in terms of the time $\tau$; the behaviour in proper time is easily obtained using (33). For D II we give the approaching behaviour, i.e. during expansion. Note that the exponents of the $\exp(\tau - \tau_*)$ function are the $\lambda_i$ (the eigenvalues of the Jacobian) listed in Tables 2 and 5. In what follows $\Omega_*$, $\Sigma_{\pm *}, \varepsilon_{\pm *}$ are arbitrary initial values for the corresponding variables: the closer these constants are chosen to the corresponding values of the variables at the stationary point, the closer is the asymptotic solution to a true trajectory falling in that point.

**Point D I** $\quad \beta = \frac{1}{2} \quad \Theta_* > 0$

$$\Omega = 1 + (\Omega_* - 1)e^{\frac{1}{3}(\tau-\tau_*)} \,, \tag{61}$$

$$\Sigma_\pm = \frac{2}{5}\left(\Sigma_{\pm *} - 3\,\varepsilon_{\pm *}\right) e^{\frac{1}{3}(\tau-\tau_*)} + \frac{3}{5}\left(\Sigma_{\pm *} + 2\,\varepsilon_{\pm *}\right) e^{-\frac{1}{2}(\tau-\tau_*)} \,, \tag{62}$$

$$\varepsilon_\pm = -\frac{1}{5}\left(\Sigma_{\pm *} - 3\,\varepsilon_{\pm *}\right) e^{\frac{1}{3}(\tau-\tau_*)} + \frac{1}{5}\left(\Sigma_{\pm *} + 2\,\varepsilon_{\pm *}\right) e^{-\frac{1}{2}(\tau-\tau_*)} \,. \tag{63}$$

Point D I represents a flat FRW universe, and the integration of the linearized system about it shows well–known behaviours. It is a 5–D saddle point that repels along the $\Omega$ axis, giving the known instability of the flat FRW model among curved ones (the *flatness problem*); then, there is a plane containing the two directions $\Sigma_\pm - 3\,\varepsilon_\pm$ (corresponding to the degeneracy of the $\lambda = \frac{1}{3}$ eigenvalue) from which D I is escaped, and a plane, containing the directions $\Sigma_\pm + 2\,\varepsilon_\pm$ (for $\lambda = -\frac{1}{2}$), from which is approached. The normals to these two planes give the directions for the purely decaying and purely growing perturbation modes about a flat matter dominated FRW universe. It is in fact straightforward to check, using (33) and $\beta = \frac{1}{2}$, that the two modes above go as $t^{\frac{2}{3}}$ and $t^{-1}$, as expected.

**Point D II** $\beta = \frac{1}{3}$ $\Theta_* > 0$

$$\Omega = \Omega_* e^{-\frac{1}{3}(\tau - \tau_*)} , \tag{64}$$

$$\Sigma_\pm = [\Sigma_{\pm *} - \varepsilon_{\pm *}(\tau - \tau_*)]\, e^{-\frac{1}{3}(\tau - \tau_*)} , \tag{65}$$

$$\varepsilon_\pm = \varepsilon_{\pm *} e^{-\frac{1}{3}(\tau - \tau_*)} . \tag{66}$$

Point D II represents a Milne universe: any volume element that falls in its basin of attraction becomes a spherical void. The fact that the fate of ever–expanding configurations is a spherical void is a well–known result of Newtonian theory (Icke 1984; see also Bertschinger 1985), and it is quite interesting that we recover the same result in the fully relativistic context.

**Point D III** $\beta = \frac{1}{2}$ $\Theta_* < 0$

$$\Omega = \Omega_* , \tag{67}$$

$$\Sigma_+ = -\frac{1}{6} + \left(\Sigma_{+*} - \varepsilon_{+*} + \frac{1}{9}\,\Omega_* + \frac{1}{6}\right) e^{-\frac{1}{2}(\tau - \tau_*)} , \tag{68}$$

$$\Sigma_- = (\Sigma_{-*} + \varepsilon_{-*}) e^{\frac{1}{2}(\tau - \tau_*)} - \varepsilon_{-*} e^{-\frac{1}{2}(\tau - \tau_*)} , \tag{69}$$

$$\varepsilon_+ = \left(\varepsilon_{+*} + \frac{1}{18}\,\Omega_*\right) e^{\frac{1}{2}(\tau - \tau_*)} - \frac{1}{18}\,\Omega_* , \tag{70}$$

$$\varepsilon_- = \varepsilon_{-*} e^{-\frac{1}{2}(\tau - \tau_*)} \tag{71}$$

This saddle point is a particular Szekeres solution, which belongs to the subset of trajectories obtained by setting $M = N = 0$ in the solutions of Section 4.3; it corresponds to Minkowsky space–time in a disguised form. In order not to escape away from it, one has to start with special initial conditions: $\Sigma_{-*} = -\varepsilon_{-*}$ and $\varepsilon_{+*} = -\frac{1}{8}\Omega_*$; further, $\Omega_* = 0$ is required to fall in D III.

**Point D IV** $\beta = 1$ $\Theta_* < 0$

$$\Omega = \Omega_* e^{-(\tau - \tau_*)} , \tag{72}$$

$$\begin{aligned}\Sigma_+ = {} & -\frac{1}{3} + \left\{\left[\left(\Sigma_{+*} + \frac{1}{3}\right) + \left(\varepsilon_{+*} + \frac{1}{18}\Omega_*\right)\right]\right. \\ & \left. + \frac{1}{18}\left(\Omega_* - \varepsilon_{+*}\right)(\tau - \tau_*)\right\} e^{-(\tau - \tau_*)} - \left(\varepsilon_{+*} + \frac{1}{18}\Omega_*\right) e^{-2(\tau - \tau_*)} ,\end{aligned} \tag{73}$$

$$\Sigma_- = \left(\Sigma_{-*} - \varepsilon_{-*}\right) e^{-(\tau - \tau_*)} + \varepsilon_{-*} , \tag{74}$$

$$\varepsilon_+ = \left(\varepsilon_{+*} + \frac{1}{18}\Omega_*\right) e^{-2(\tau - \tau_*)} - \frac{1}{18}\Omega_* e^{-(\tau - \tau_*)} , \tag{75}$$

$$\varepsilon_- = \varepsilon_{-*} . \tag{76}$$

This point represents the degenerate (Szekeres) pancake. Starting with generic initial conditions in its neighborhood, the trajectory ends up in a nearby triaxial point of the family T III, unless the special initial condition $\varepsilon_{-*} = 0$ is chosen, for which the trajectory falls onto D IV. This result, as well as the asymptotic behaviour of $\varepsilon_\pm$ near the singularity, is in complete agreement with the numerical analysis by Croudace et al. (1994), who first noticed the instability of the planar pancake against general (i.e. non–degenerate) initial perturbations of the shear and electric tidal field. Taking $\varepsilon_{-*}$ as small expansion parameter, the nearby T III point is characterized by $\Sigma_+ \simeq -\frac{1}{3} + \frac{1}{2}\varepsilon_{-*}^2$ and $\varepsilon_+ \simeq \frac{1}{2}\varepsilon_{-*}^2$, while $\Sigma_- \simeq \varepsilon_-\varepsilon_{-*}$, thus at first order in $\varepsilon_{-*}$ the difference between D IV and the nearby T III point is seen only in $\Sigma_-$ and $\varepsilon_-$, as shown above.

**Point D V** $\quad \beta = 1 \quad \Theta_* < 0$

$$\Omega = \Omega_* e^{-(\tau-\tau_*)} , \tag{77}$$

$$\Sigma_+ = \frac{1}{3} - \frac{1}{6}\Omega_* e^{-(\tau-\tau_*)} + \left(\varepsilon_{+*} - \frac{2}{3}\Sigma_{+*} + \frac{1}{18}\Omega_*\right) e^{-\frac{2}{3}(\tau-\tau_*)} \tag{78}$$

$$+ \left(\frac{5}{3}\Sigma_{+*} - \varepsilon_{+*} - \frac{1}{3} + \frac{1}{9}\Omega_*\right) e^{-\frac{5}{3}(\tau-\tau_*)} , \tag{79}$$

$$\Sigma_- = \frac{2}{3}(2\Sigma_{-*} + \varepsilon_{-*}) - \frac{1}{5}(\Sigma_{-*} + 3\,\varepsilon_{-*}) e^{-\frac{5}{3}(\tau-\tau_*)} , \tag{80}$$

$$\varepsilon_+ = \frac{2}{9} - \frac{1}{6}\Omega_* e^{-(\tau-\tau_*)} + \frac{5}{3}\left(\varepsilon_{+*} - \frac{2}{3}\Sigma_{+*} + \frac{1}{18}\Omega_*\right) e^{-\frac{2}{3}(\tau-\tau_*)} \tag{81}$$

$$+ \left(\frac{5}{3}\Sigma_{+*} - \varepsilon_{+*} - \frac{1}{3} + \frac{1}{9}\Omega_*\right) e^{-\frac{5}{3}(\tau-\tau_*)} , \tag{82}$$

$$\varepsilon_- = -\frac{1}{5}(2\,\Sigma_{-*} + \varepsilon_{-*}) + \frac{2}{5}(\Sigma_{-*} + 3\,\varepsilon_{-*}) e^{-\frac{5}{3}(\tau-\tau_*)} . \tag{83}$$

This point represents the degenerate spindle. The behaviour in its neighborhood is very similar to that around D IV. The generic trajectory falls in a nearby point T III, at first order seen only in $\Sigma_-$ and $\varepsilon_-$, and characterized by $\Sigma_{-*} = -3\,\varepsilon_{-*}$. The special initial condition required to fall onto D V is $\varepsilon_{-*} = -2\,\Sigma_{-*}$, explicitly showing that this point is an attractor even for (a set of measure zero of) triaxial configurations.

**Point D VI** $\quad \beta = \frac{3}{8} \quad \Theta_* > 0$

$$\Omega = \Omega_* e^{-\frac{1}{4}(\tau-\tau_*)} , \tag{84}$$

$$\Sigma_+ = -\frac{1}{12} + \left(\frac{5}{7}\Sigma_{+*} + \frac{8}{7}\varepsilon_{+*} - \frac{23}{63}\Omega_* + \frac{1}{42}\right) e^{-\frac{5}{8}(\tau-\tau_*)} \tag{85}$$

$$+ \left(\frac{2}{7}\Sigma_{+*} - \frac{8}{7}\varepsilon_{+*} - \frac{19}{63}\Omega_* + \frac{5}{84}\right) e^{\frac{1}{4}(\tau-\tau_*)} + \frac{2}{3}\Omega_* e^{-\frac{1}{4}(\tau-\tau_*)} , \tag{86}$$

$$\Sigma_- = \left[\Sigma_{-*}\cos\left(\frac{\sqrt{15}}{16}(\tau-\tau_*)\right) - \frac{1}{\sqrt{15}}\left(16\,\varepsilon_{-*} - 3\,\Sigma_{-*}\right)\sin\left(\frac{\sqrt{15}}{16}(\tau-\tau_*)\right)\right] e^{-\frac{5}{16}(\tau-\tau_*)} \quad (87)$$

$$\varepsilon_+ = \frac{1}{32} + \left(\frac{5}{28}\,\Sigma_{+*} + \frac{2}{7}\,\varepsilon_{+*} - \frac{23}{252}\,\Omega_* + \frac{1}{168}\right) e^{-\frac{5}{8}(\tau-\tau_*)} \quad (88)$$

$$+ \left(-\frac{5}{28}\,\Sigma_{+*} + \frac{5}{7}\,\varepsilon_{+*} + \frac{95}{504}\,\Omega_* - \frac{25}{672}\right) e^{\frac{1}{4}(\tau-\tau_*)} - \frac{7}{72}\,\Omega_* e^{-\frac{1}{4}(\tau-\tau_*)}\,, \quad (89)$$

$$\varepsilon_- = \left[\varepsilon_{-*}\cos\left(\frac{\sqrt{15}}{16}(\tau-\tau_*)\right) - \frac{1}{\sqrt{15}}\left(3\,\varepsilon_{-*} - \frac{3}{2}\,\Sigma_{-*}\right)\sin\left(\frac{\sqrt{15}}{16}(\tau-\tau_*)\right)\right] e^{-\frac{5}{16}(\tau-\tau_*)} \quad (90)$$

This saddle point is also a particular Szekeres solution, which belongs to the subset of trajectories obtained by setting $M = 0$ in the solutions of Section 4.3. It corresponds to a particular model obtained by Bonnor & Tomimura (1976) as the $t \to \infty$ limit of a special set (PI in their paper) of ever–expanding Szekeres models. It is indeed seen from (85) and (88) above that there are special initial conditions for which the only growing mode, given by $\lambda_2 = \frac{1}{4}$ (in Table 5 the signs are given for $\Theta < 0$), is suppressed: for this special set, therefore, D VI is an attractor.

# 6 Conclusions

In this work we have carried out a detailed study of the non–linear dynamics of irrotational dust with vanishing magnetic tidal field. This type of dynamics is completely described by six first–order quasi–linear ordinary differential equations to which the set of partial differential equations given by Ellis (1971) reduces in the case $p = \omega_{ab} = H_{ab} = 0$. From the point of view of the theory of partial differential equations this means that, under the above dynamical restrictions, the initial value problem for the general equations is automatically reduced to the characteristic initial value problem, where the only surviving characteristics are the fluid flow–lines. Thus, for these models all the information regarding the environment of each fluid element is that coded in the data on the initial time surface: no sound or gravitational waves are allowed to exchange information between neighboring fluid elements after the initial time, so that these models may be termed silent universes. However, coded in the initial value for $E_{ab}$ at each fluid element location, there is always an interaction, even beyond the scale of the particle horizon, given by the Coulombian (type D) field generated by distant matter: this interaction is ultimately responsible for the spindle–like nature of the singularities we have found. The difference with Newtonian cosmology is precisely in this, i.e. in the arbitrariness of the boundary conditions for $E_{ab}$ in the Newtonian case (Ellis 1971; Zel'dovich & Novikov 1983). With appropriate boundary conditions, one can always find a Newtonian solution corresponding to the relativistic one, but the vice versa is not necessarily true (Ellis 1971, Matarrese, Pantano, & Saez 1994b).

Although the problem we have studied here is by itself an important and fascinating application of relativistic cosmology, it is tempting to try to interpret our silent universe models as an approximation, useful for more general situations, i.e. beyond those particular cases which exactly solve all the constraints required by $H_{ab} = 0$. In particular, one would be interested to know if such a formalism can be applied to the non–linear evolution of scalar perturbations in a matter–dominated FRW universe, i.e. to the general problem of gravitational instability of collisionless matter (e.g. Sahni & Coles 1994, for an updated review on the subject). This was indeed the question originally posed and left open by Matarrese, Pantano, & Saez (1993).

The obvious question *"why should one resort to general relativity in dealing with structure formation in the universe?"*, can be answered as follows. It is well–known that the standard Newtonian approximation has two main limitations (e.g. Peebles 1980): it cannot be applied on scales comparable or larger than the universe horizon size, and it cannot be applied during the phases of highly non–linear collapse of a perturbation, when the gravitational interaction becomes too strong and/or relativistic motions are produced. These two limitations of Newtonian theory lead us to speculate on two possible applications of our formalism: the description of our universe on super–horizon scales, and the study of the highly non–linear collapse of cosmological perturbations.

The first application is rather obvious. On scales larger than the Hubble radius, no causal communication is possible, so each patch of the universe on that scale is expected to evolve independently of the surrounding ones, gravitational radiation and sound waves are simply too slow to carry any useful signal. This can be understood as the "ultra–relativistic" limit of Einstein equations. This was in fact proven by Matarrese, Pantano, & Saez (1994a,b) at second order in perturbation theory, but is very likely to be valid at any order. What emerges is a new picture of the universe on ultra–large scales, with large patches of the universe evolving non–linearly according to our equations: only a small set among these will be approximately FRW, while most of them will either expand forever to end up in a spherical void, or undergo collapse toward some Kasner–like configuration.

The second, more speculative application of the present formalism is related to the late phases of expansion, or collapse after turn–around, of scalar perturbations in a FRW background. As far as the fate of ever–expanding configurations is concerned, we think that the situation is almost completely settled down: our attractor for the expansion phase (D II) is the classical spherical void, already discovered in the Newtonian context (Icke 1984; see also Bertschinger 1985, Bertschinger & Jain 1994). That during the late, free expansion of an underdense region the surrounding matter has no dynamical effect should cause no surprise: in such a case the gravitational field does not grow too large, while the overall tendency to make the perturbation spherical prevents the occurrence of a relevant magnetic component. In such a case the Newtonian approximation already gives the right answer to the problem.

More problematic is the dynamics of collapse. The present status of our understanding of the problem leads to the following picture of the history of a scalar perturbation. At early times the magnetic component is negligible (Goode 1989; Bruni, Dunsby, & Ellis 1992), as it only appears as a second order effect (Matarrese, Pantano, & Saez 1994a,b). Later on, however, its presence is generally caused by the spatial gradients of the initial perturbation field. At this, mildly non–linear stage, the presence of $H_{ab}$ signals a non–zero flux of gravitational information coming from the environment: it tells to the electric tidal field that the state of the surrounding matter has changed compared to the initial conditions; as a consequence the electric tide modifies the shear field, which, in turn, acts on the fluid expansion and density, as prescribed by the Raychaudhuri and mass conservation equations. From this moment onwards the evolution of a fluid element is a mixture of two competing effects: that determined by the imprint of its "private" initial conditions, and that arising later from the influence of the "rest of the world". Which of these two effects will be dominant depends on many variables: the form of the particular fluid element, the nature of the initial conditions, such as the type of collisionless matter, the statistics of primordial perturbations, and so on. For instance, a recent study within Newtonian theory (Eisenstein & Loeb 1994) shows that, at least in the case of the standard cold dark matter model, the evolution of shapes of fluid elements is primarily induced by the external tide, and not by their initial triaxiality.

What we believe still to be an open problem is what happens much after a fluid element has turned around and its evolution has detached from the overall expansion of the universe. Maybe, at this stage, neglecting the influence of the surrounding matter, and thus $H_{ab}$ (in the absence of pressure gradients), becomes more feasible. In that case, the dynamics we have discussed so far should apply, and the final fate of the fluid element should be that of ending up into some member of the triaxial attractor set for collapse (T III and $\overline{\text{T III}}$), i.e. in some Kasner–like singularity.

Notice that we are not claiming that there is a one–to–one mapping between the particular final configuration and the specific initial condition of the collapsing fluid element, because this would disregard the influence of the environment in between. Rather, the picture we have in mind is that of fluid elements starting their evolution and ending it according to our local set of equations, but evolving in the middle in a much more complicated and non–local way. There is an important corollary to our conjecture: since the pancake configuration (D IV and its replicas) only attracts exactly degenerate fluid elements, no fluid element (actually only a set of measure zero) would undergo pancake collapse, i.e. no shell–crossing will ever occur with non–degenerate initial data: the fate of every collapsing element in a pressureless fluid should be a triaxial spindle singularity.

However, if the latter conjecture should prove incorrect, then we should wonder in which sense a non–zero $H_{ab}$ component might modify the above picture. We may try to make some guess, this problem being related to the more general one of the robustness (Coley & Tavakol 1992) of the $p = \omega_{ab} = H_{ab}- = 0$ models. First of all, let us remind that all our stationary points are also special exact solutions of the most general system, which is obtained by accounting for the magnetic Weyl tensor and for non–zero pressure. Then, the magnetic tensor can only act as follows: *i)* it can introduce new non–trivial stationary configurations, *ii)* it can stabilize some of our saddle points or repellers, *iii)* it can destabilize some of our collapse attractors, and *iv)* it can modify the basin of attraction for some of our stable points, e.g. it can act in such a way that the pancake singularity also attracts a subset of triaxial trajectories. All these possibilities, except case *i)*, can be studied by considering small perturbations around our stationary points, with $H_{ab}$ switched on. In some cases we already know the answer: for instance, the standard theory of linear perturbations around FRW tells us that point D I is not stabilized by tensor perturbations (i.e. by a linear contribution from the $H_{ab}$ component). The remaining cases need to be studied, and this will be the aim of our future investigation. In any case, there is one conclusion we can anticipate: if one had to find that the basin of attraction of our planar pancake (point D IV and its replicas) is enlarged by the presence of a non–zero $H_{ab}$ component it would just mean that the shell–crossing singularity of Newtonian theory (e.g. Shandarin & Zel'dovich 1989) only occurs in the non–degenerate relativistic problem because of the non–linear interaction with the environment: a non trivial conclusion indeed!

Let us conclude with an overview of few more open questions.

First, we have attacked here the problem of the time evolution in silent universes, assuming that suitable initial data exist that satisfy the constraint equations for these models (see Barnes & Rowlingson 1989). Certainly there exists a subset of perturbed FRW initial conditions that satisfy these constraints exactly: it will therefore be interesting to further study them, in order to see how much restricted they are. More generally, it will be enlightening to completely solve the equations for silent universes, generalizing Szekeres ones, as these solutions will very likely reveal further interesting features of these models, as well as some *intrinsic symmetry* they must possess (cf. Kramer et al. 1980, MacCallum 1993).

We have shown that the final fate of generic triaxial configurations that turn around is to recollapse to one of our spindle singularities of the Kasner attractor. Related to this, there are at least two open questions we want to point out, namely: if these singularities are real, and, being real, if they are naked. The reality of singularities occurring in the collapse of a fluid of dust is a controversial issue, essentially because of the lack of any pressure gradient working against gravity (e.g. Rendall 1992). On the other hand, what really matters from the point of view of cosmological structure formation is not whether the curvature singularities still occur in the presence of pressure gradients, but if the triaxial collapsing configurations we have found give any answer to the question if filamentary or pancake–like structures are preferred. Regarding the second point, this clearly goes beyond the scopes of the present work. Further, the "time reversal" property of our formulation of the dynamical problem for silent universes shows that the Kasner set is also the repeller for generic expanding trajectories, so that the typical initial singularity is also spindle–like. Related to this are the issues of the isotropy of the observed universe, of its initial state, and of the arrow of time. In silent universes, as well as in the Szekeres subcase, there are trajectories that spring off point D I, representing the Einstein–de Sitter model, showing that it is conceivable to think of an initial isotropic singularity: this will be preferred in a *quiescent cosmology*, according to which the universe was initially highly regular, this fact in turn being motived by Penrose's idea of a link between gravitational entropy and the Weyl tensor, selecting the arrow of time (see Goode & Wainwright 1992, and reference therein).

Of course, an alternative to this point of view is provided by the inflationary scenario. We will show in a forthcoming article (Bruni, Matarrese, & Pantano 1994) that patches of a silent universe with non–zero cosmological constant that do not recollapse expand toward a locally de Sitter universe.

# Acknowledgements

J. Miller is acknowledged for useful comments and discussions. MB would like to thank British PPARC (grant GR/J 36440) and Università degli Studi di Trieste for financial support, and SISSA for hospitality during the preparation of this work. SM and OP acknowledge Italian

MURST for financial support. We have used Maple V to solve for the stationary points of our systems of differential equations, and to compute the eigenvalues of the Jacobian of these systems.

# A Parametric representation

In this paper we have used the variables $\Sigma_\pm$ and $\varepsilon_\pm$ to describe the shear of the fluid flow and the electric tidal field. As it was explained in Section 3 this choice is very convenient in order to select the subsystem describing the degenerate case $\Sigma_- = \varepsilon_- = 0$ (the Szekeres models), and also in order to obtain the polynomial character of the flux $\vec{F}(\vec{G})$ in Eq.(23) [the right hand side on the system (26)–(30)] which has been used to find the stationary points of the system. This choice, however, somehow privileges the third axis (or the 1–2 plane orthogonal to it). Another possible choice of variable is that used by Bertschinger & Jain (1994), which is based on using the magnitude of the shear $\sigma \equiv \sqrt{\frac{1}{2}\sigma_{ab}\sigma^{ab}}$ and of the electric tidal field $E = \sqrt{\frac{1}{2}E_{ab}E^{ab}}$, and two angles $\alpha$ and $\beta$. We have used these variables only in the parametric presentation of the attractor. Here we give the basic definitions, and the relations to the work of Bertschinger & Jain (1994). Actually, the quantities used by Bertschinger & Jain (1994) differ by ours, in that they define the velocity field using the conformal time, rather than the proper time; this difference is however irrelevant to the present discussion.

Bertschinger & Jain (1994) represent the shear by $\sigma_{ab} = \frac{2}{3}\sigma_{BJ}Q_{ij}$, with $\sigma_{BJ} < 0$ and

$$Q_{ij} = \mathrm{diag}\left[\cos(\tfrac{\alpha+2\pi}{3}),\, \cos(\tfrac{\alpha-2\pi}{3}),\, \cos(\tfrac{\alpha}{3})\right].$$

One has $Q^2 = \frac{3}{2}$, giving $\sigma^2 = \frac{1}{2}(\sigma_1^2 + \sigma_2^2 + \sigma_3^2) = \sigma_1^2 + \sigma_2^2 + \sigma_1\sigma_2 = 3\sigma_+{}^2 + \sigma_-{}^2 = \frac{1}{3}\sigma_{BJ}^2$, i.e. $\sigma_{BJ} = -\sqrt{3}\sigma$. Similar formulas were given for $E_{ab}$. In (Bertschinger & Jain 1994) it was found that $\sigma_{BJ} = \theta \to -\infty$ at the end of the collapse (with $\theta$ the peculiar expansion scalar), corresponding to our exact result $\Sigma = \frac{\sigma}{|\Theta|} = \frac{1}{3}$ (the background FRW part of $\Theta$ is completely negligible in the phase of advanced collapse, so $\Theta = \theta$ at this stage).

In this work we have used $\alpha = \frac{\alpha_{BJ}}{3} - \pi$, and $\beta = \frac{\beta_{BJ}}{3}$, so that the relation $\alpha_{BJ} + 2\beta_{BJ} = 3\pi$ goes over $\beta = -\frac{\alpha}{2}$. The angle $\alpha$ and $\beta$ are the angles in the planes $\{\Sigma_+, \frac{1}{\sqrt{3}}\Sigma_-\}$ and $\{\varepsilon_+, \frac{1}{\sqrt{3}}\varepsilon_-\}$ respectively (measured counterclockwise), and are used for the parametric plots in Figure 4.

Using $\Sigma^2 = 3\Sigma_+{}^2 + \Sigma_-{}^2$ and $\varepsilon^2 = 3\varepsilon_+{}^2 + \varepsilon_-{}^2$, ($\varepsilon = \frac{E}{\Theta^2}$) the relations between $\Sigma_\pm$ and $\Sigma$, $\alpha$ and $\varepsilon_\pm$ and $\varepsilon$, $\beta$ are :

$$\begin{aligned} \Sigma_+ &= \frac{1}{\sqrt{3}}\Sigma(\alpha)\cos(\alpha)\,, & \varepsilon_+ &= \frac{1}{\sqrt{3}}\varepsilon(\beta)\cos(\beta)\,, \\ \Sigma_- &= \Sigma(\alpha)\sin(\alpha)\,, & \varepsilon_- &= \varepsilon(\beta)\sin(\beta)\,, \end{aligned} \tag{91}$$

with $\Sigma$ and $\varepsilon$ functions of the point.

# Figure captions

**Figure 1:** The phase–space flow for Szekeres models during collapse $\Theta < 0$, above points D II and D V, placed at the center of the bottom plane $\Omega = 0$ of the boxes; arrows length is rescaled in each case (Om is $\Omega$, ep is $\varepsilon_+$, and Sp is $\Sigma_+$). From top to bottom, on the left: a) the flow above point D II up to $\Omega \approx 2$ and including point D I, edge view; b) top view of the flow above point D I; c) edge view of the flow above point D II. On the right (top to bottom): a) the edge view of the flow above point D V up to $\Omega \approx 2$; b) top view above D V; c) edge view.

**Figure 2:** The $\Omega = 0$ (vacuum) plane for Szekeres model. Top: the Poincaré disk of the plane $\Sigma_+$, $\varepsilon_+$ (conformal mapping of it) is shown for $\Theta < 0$. Bottom: the central region of the plane $\Sigma_+$, $\varepsilon_+$ including all the five stationary points is shown for $\Theta > 0$. The trajectories are the same, but the directions are reversed: the two stable nodes above are unstable for $\Theta > 0$, while the origin is now stable; it represents the final fate of voids. Note that the $\varepsilon_+ = 0$ axis is a separatrix between the $\varepsilon_+ > 0$ and the $\varepsilon_+ < 0$ worlds; since $H_{ab} = 0$ by assumption, and $\rho = 0$ on the plane in figure, the $\varepsilon_+ = 0$ line represents Minkowski spacetime: in particular the three points on the line from left to right are D IV, D II, and D III. The two points above the $\varepsilon_+ = 0$ axis are D VI (left) and D V (right).

**Figure 3:** The Poincaré disk of the plane $\Theta$, $\sigma_+$ corresponding the the $\varepsilon_+ = 0$ axis in Figure 2. The behaviour of these Minkowski trajectories simply illustrate that of more complicated situations. The central stationary point is Minkowski spacetime in its usual form (static and shear free). In these conformal maps straight lines stay straight: the vertical $\sigma_+ = 0$ axis is Milne universe, point D II (expanding in the upper half, contracting in the lower half); the two straight lines $\Theta = 6\sigma_+$ and $\Theta = -3\sigma_+$ correspond to points D III and DIV respectively.

**Figure 4:** The whole attractor family $\{\mathrm{T\,III}, \overline{\mathrm{T\,III}}\}$ as it shows up in the $\{\Sigma_+, \Sigma_-\}$ plane (upper left) and in the $\{\varepsilon_+, \varepsilon_-\}$ plane (upper right). The dashed lines are $\Sigma_-/\Sigma_+ = 0, \pm 3$ and $\varepsilon_-/\varepsilon_+ = 0, \pm 3$, corresponding to degenerate configurations, and divide the planes in six physically equivalent sectors. Dots represent degenerate points of the attractor. The spindle is the $\Sigma_+ = \frac{1}{3}$ dots, with its two replicas at $\Sigma_+ = -\frac{1}{6}$; the other three points are the pancake. Since this latter is conformally flat, it appears at the origin of the $\{\varepsilon_+, \varepsilon_-\}$ plane, where the three replicas coincide; the other three points in this plane are the spindle. Lower figures show $\varepsilon_+ = \varepsilon_+(\Sigma_+)$ and $\varepsilon_- = \varepsilon_-(\Sigma_-)$.

Table 1: Stationary points for degenerate (D) models

| Point | $\Omega$ | $\Sigma_+$ | $\varepsilon_+$ | Collapse | Expansion | Model |
|---|---|---|---|---|---|---|
| D I | 1 | 0 | 0 | *spherical* | *spherical* | *Flat FRW* |
| D II | 0 | 0 | 0 | *spherical* | *spherical* | *Milne* (M) |
| D III | 0 | 1/6 | 0 | *prolate* | *oblate* | *Szekeres* (M) |
| D IV | 0 | $-1/3$ | 0 | *oblate* | *prolate* | *Kasner* (M) |
| D V | 0 | 1/3 | 2/9 | *prolate* | *oblate* | *Kasner* |
| D VI | 0 | $-1/12$ | 1/32 | *oblate* | *prolate* | *Szekeres* |
| D VII | $-3$ | $-1/3$ | 1/6 | – | – | *unphysical* |

Table 2: Eigenvalues of the Jacobian $J(\vec{g}_S, \Theta < 0)$ for degenerate (D) models, and point type for $\Theta < 0$ and $\Theta > 0$

| Point | $\lambda_1$ | $\lambda_2$ | $\lambda_3$ | Collapse | Expansion |
|---|---|---|---|---|---|
| D I | $-1/3$ | $-1/3$ | 1/2 | *saddle* | *saddle* |
| D II | 1/3 | 1/3 | 1/3 | *repeller* | *attractor* |
| D III | 0 | $-1/2$ | 1/2 | *saddle* | *saddle* |
| D IV | $-1$ | $-1$ | $-2$ | *attractor* | *repeller* |
| D V | $-1$ | $-5/3$ | $-2/3$ | *attractor* | *repeller* |
| D VI | 1/4 | $-1/4$ | 5/8 | *saddle* | *saddle* |
| D VII | $-1$ | $-1/2$ | 1 | *saddle* | *saddle* |

Table 3: Exponents and singularity type for the stationary solutions

| Point | $\beta$ | $p_1$ | $p_2$ | $p_3$ | Singularity |
|---|---|---|---|---|---|
| D I | 1/2 | 2/3 | 2/3 | 2/3 | *point-like* |
| D II | 1/3 | 1 | 1 | 1 | *point-like* |
| D III | 1/2 | 1 | 1 | 0 | *cylinder* |
| D IV | 1 | 0 | 0 | 1 | *pancake* |
| D V | 1 | 2/3 | 2/3 | $-1/3$ | *spindle* |
| D VI | 3/8 | 2/3 | 2/3 | 4/3 | *point-like* |
| T III | 1 | $\sum p_\alpha = \sum p_\alpha^2 = 1$ | | | *spindle or pancake* |

Table 4: Stationary points for the general triaxial (T) system

| Point | $\Omega$ | $\Sigma_+$ | $\Sigma_-$ | $\varepsilon_+$ | $\varepsilon_-$ | Collapse | Expansion | Model |
|---|---|---|---|---|---|---|---|---|
| T I | 0 | $-1/12$ | 1/4 | 0 | 0 | *elongated* | *flattened* | *Szekeres* (M) |
| T II | 0 | 1/24 | 1/8 | $-1/64$ | $-3/64$ | *flattened* | *elongated* | *Szekeres* |
| T III | 0 | $\Sigma_+$ | $\Sigma_-(\Sigma_+)$ | $\varepsilon_+(\Sigma_+)$ | $\varepsilon_-(\Sigma_+)$ | *flattened or elongated* | | *Kasner* |
| T IV | $-3$ | 1/6 | 1/2 | $-1/12$ | $-1/4$ | – | – | *unphysical* |

Table 5: Eigenvalues of the Jacobian $J(vecG_S, \Theta < 0)$ for the general triaxial (T) models, and point type

| Point | $\lambda_1$ | $\lambda_2$ | $\lambda_3$ | $\lambda_4$ | $\lambda_5$ | Collapse | Expansion |
|---|---|---|---|---|---|---|---|
| D I | $-1/3$ | $1/2$ | $1/2$ | $-1/3$ | $-1/3$ | *saddle* | *saddle* |
| D II | $1/3$ | $1/3$ | $1/3$ | $1/3$ | $1/3$ | *repeller* | *attractor* |
| D III | $0$ | $-1/2$ | $-1/2$ | $1/2$ | $1/2$ | *saddle* | *saddle* |
| D IV | $-1$ | $0$ | $-2$ | $-1$ | $-1$ | *attractor* | *repeller* |
| D V | $-1$ | $0$ | $-2/3$ | $-5/3$ | $-5/3$ | *attractor* | *repeller* |
| D VI | $1/4$ | $-1/4$ | $5/8$ | $5/16 - i\sqrt{15}/16$ | $5/16 + i\sqrt{15}/16$ | *saddle* | *saddle* |
| D VII | $-1$ | $-1/2$ | $1$ | $1/4 - i7/4$ | $1/4 + i7/4$ | *saddle* | *saddle* |
| T I | $0$ | $-1/2$ | $-1/2$ | $1/2$ | $1/2$ | *saddle* | *saddle* |
| T II | $1/4$ | $-1/4$ | $5/8$ | $5/16 - i\sqrt{15}/16$ | $5/16 + i\sqrt{15}/16$ | *saddle* | *saddle* |
| T III | $-1$ | $0$ | $-4/3 + 2\Sigma_+$ | $-4/3 - \Sigma_+ - \sqrt{1 - 9{\Sigma_+}^2}/\sqrt{3}$ | $-4/3 - \Sigma_+ + \sqrt{1 - 9{\Sigma_+}^2}/\sqrt{3}$ | *attractor* | *repeller* |
| T IV | $-1$ | $-1/2$ | $1$ | $1/4 - i\sqrt{7}/4$ | $1/4 + i\sqrt{7}/4$ | *saddle* | *saddle* |

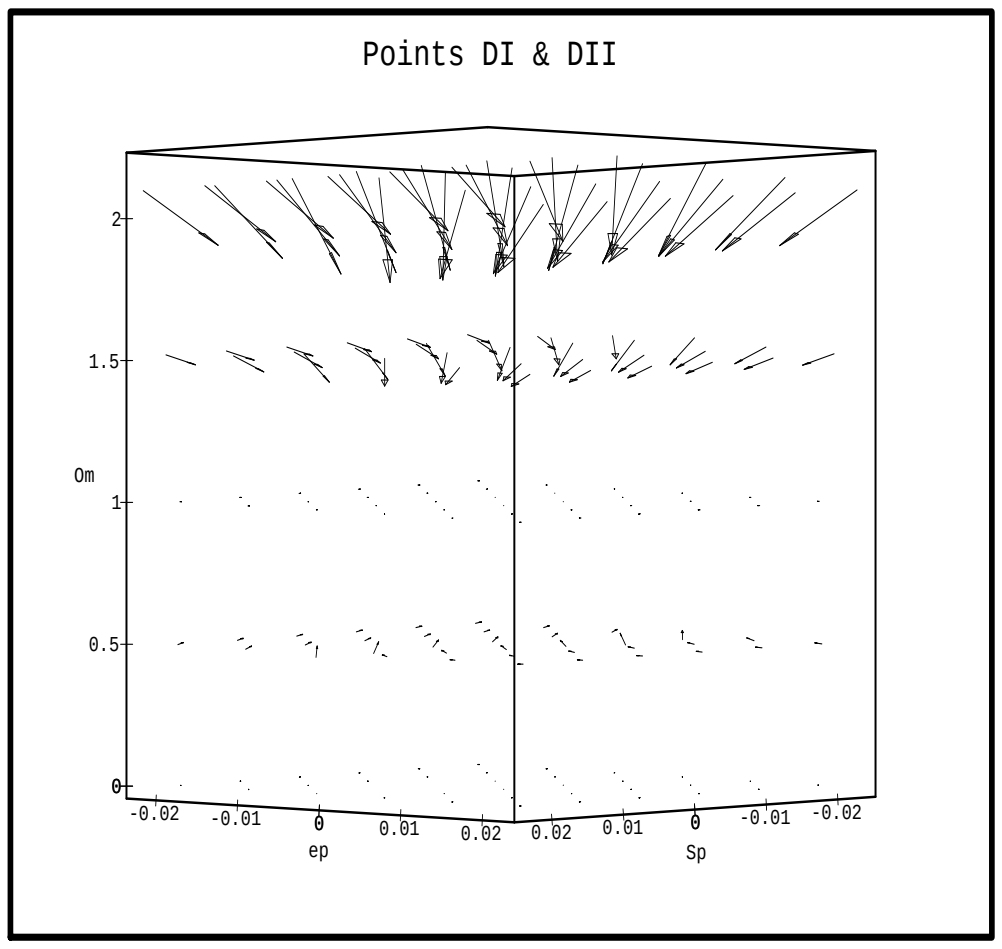
Points DI & DII
Om
ep
Sp

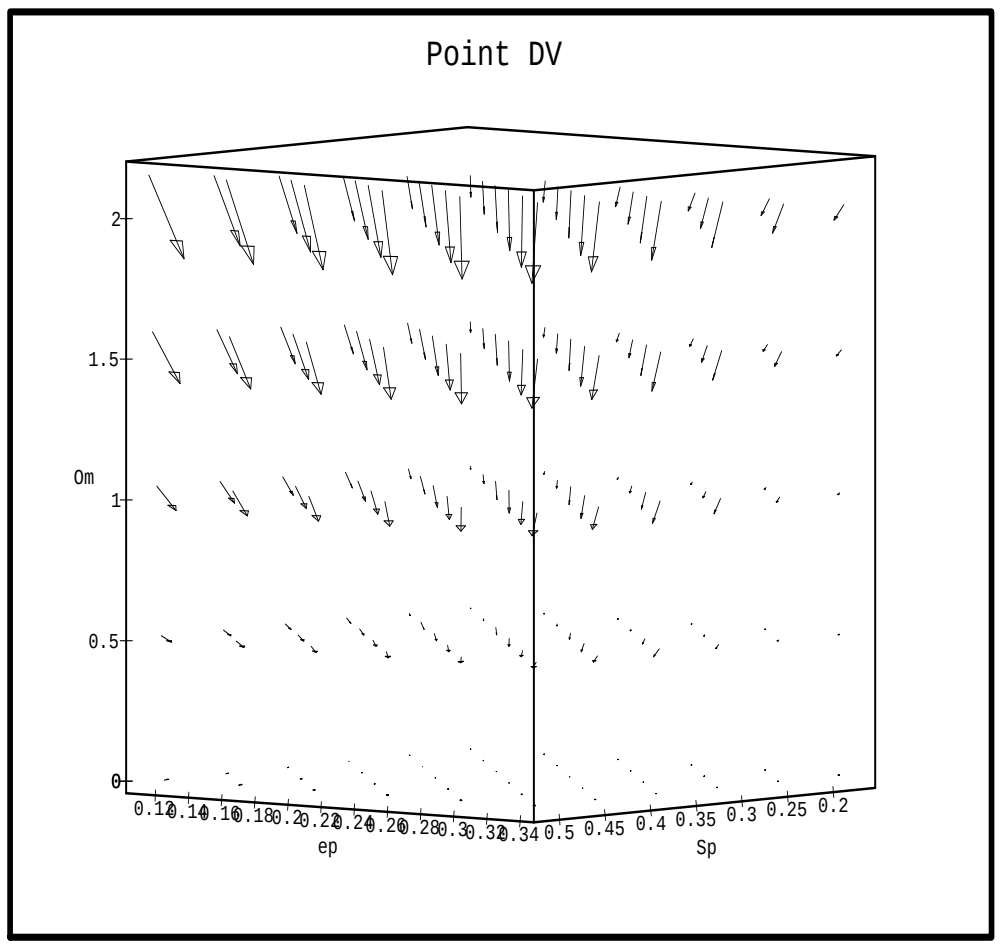
Point DV
Om
ep
Sp

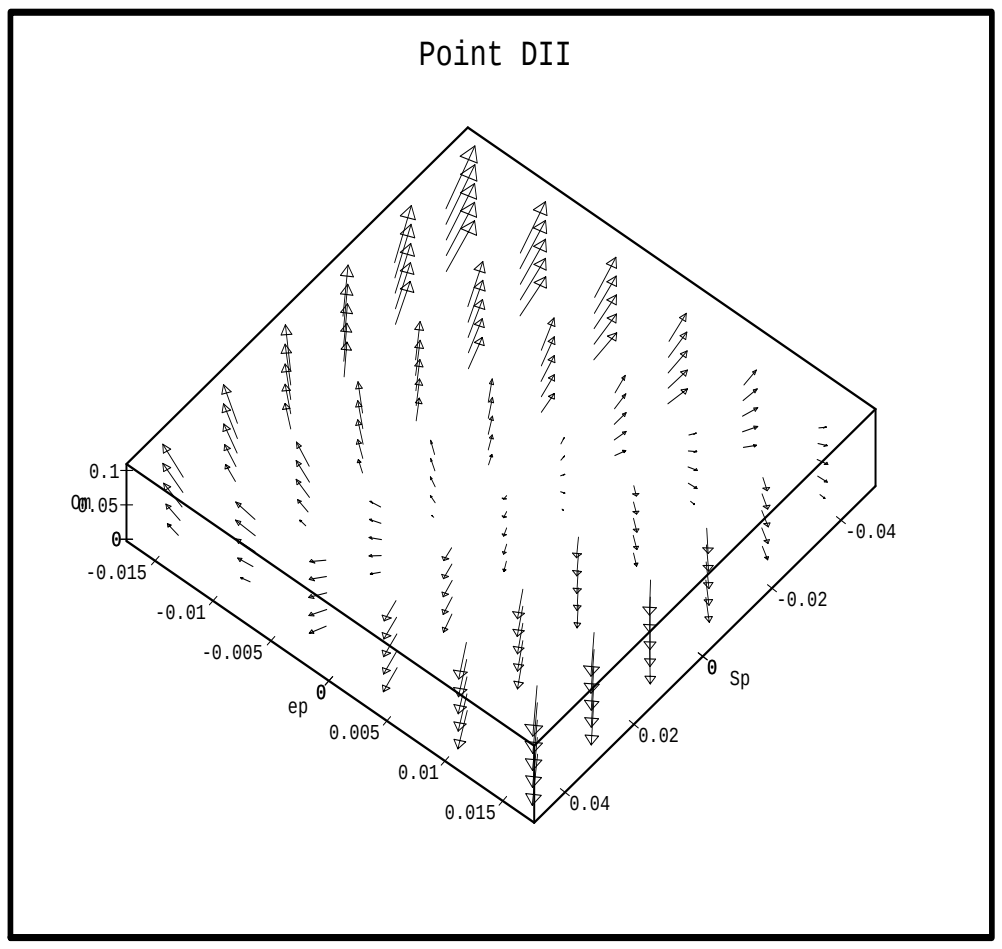
Point DII
Om
ep
Sp

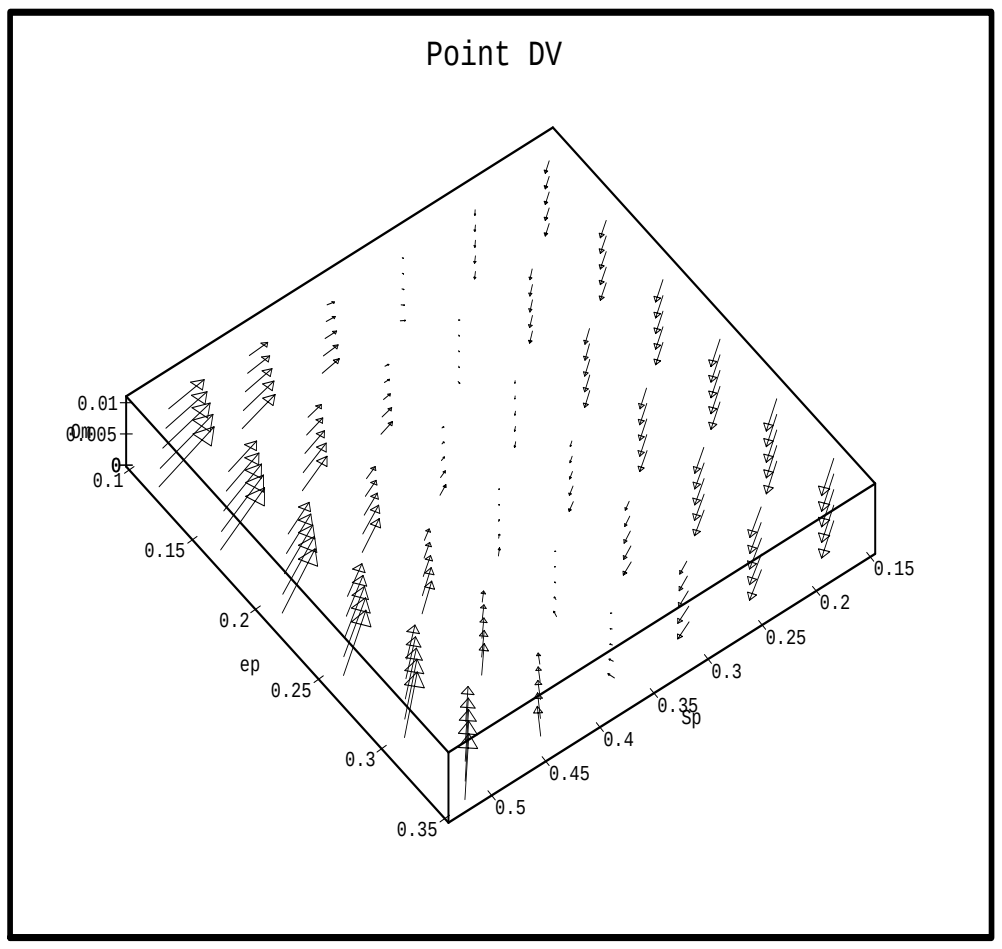
Point DV
Om
ep
Sp

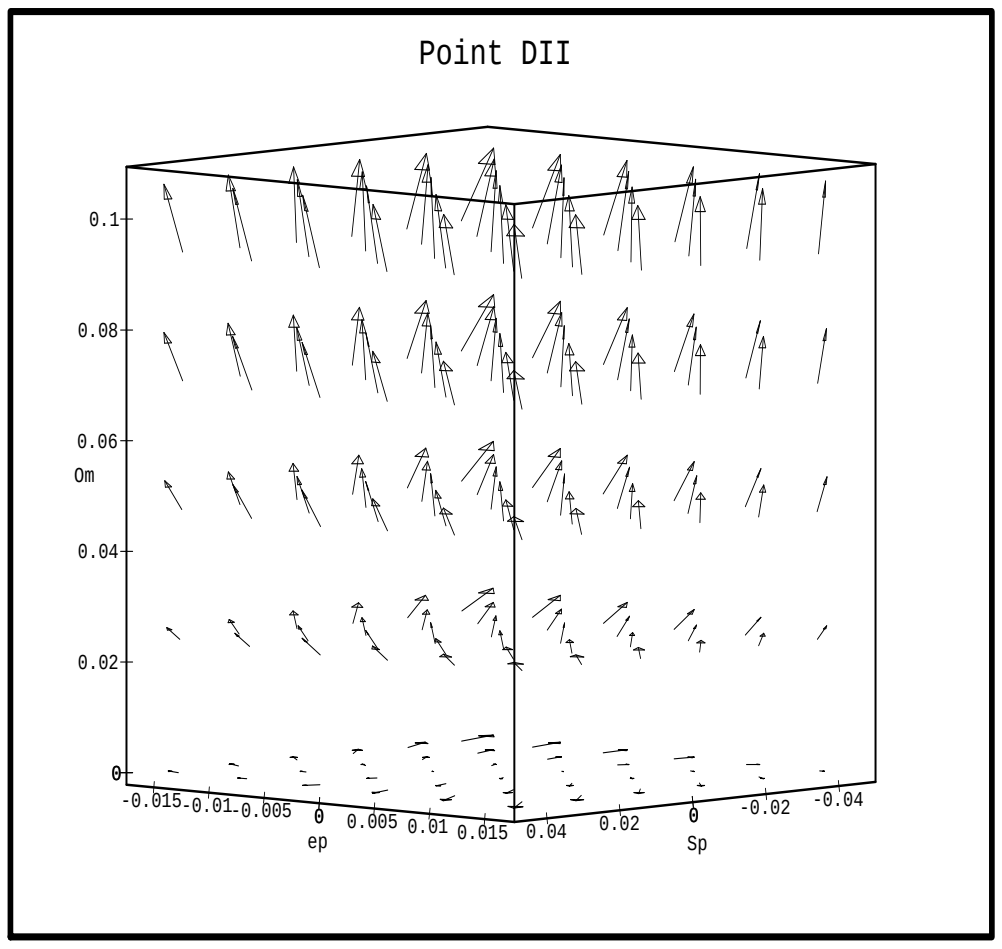
Point DII
Om
ep
Sp

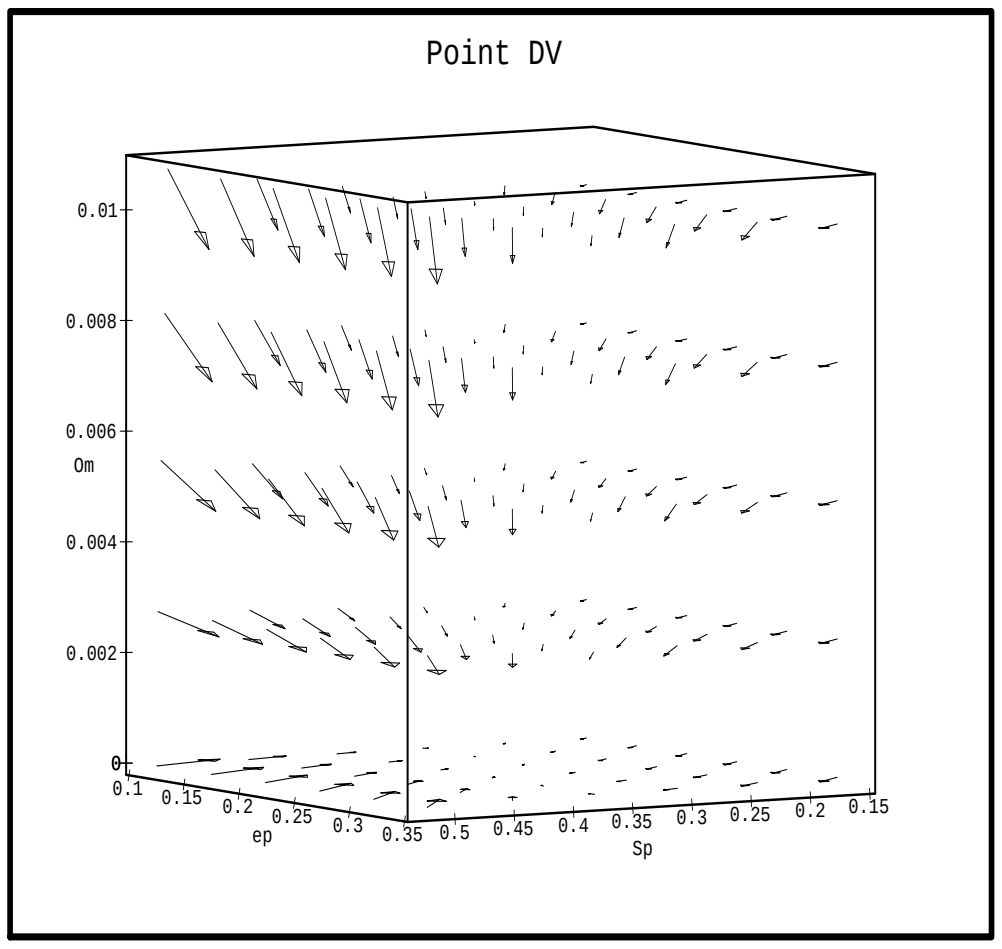
Point DV
Om
ep
Sp

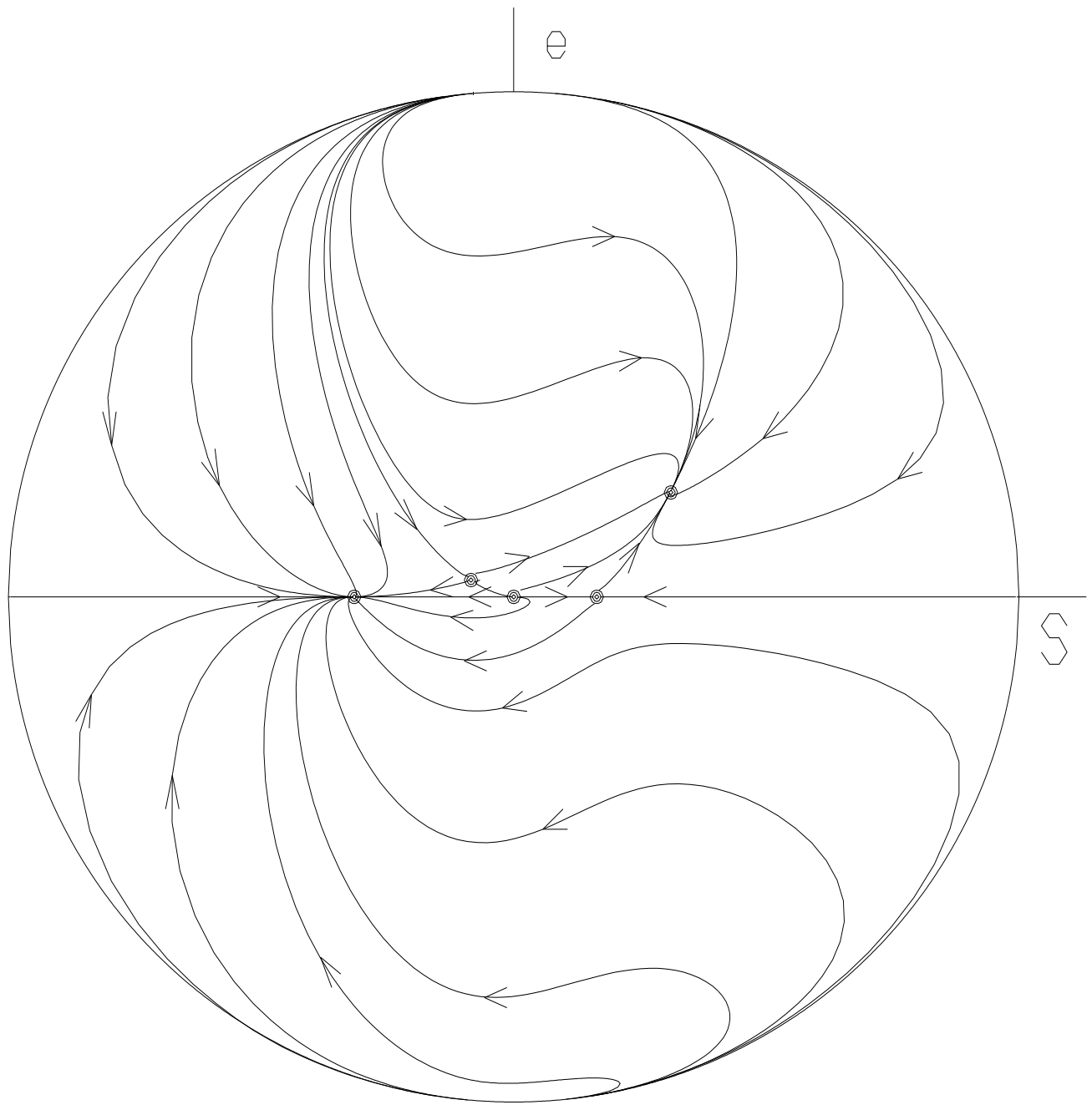
e
S

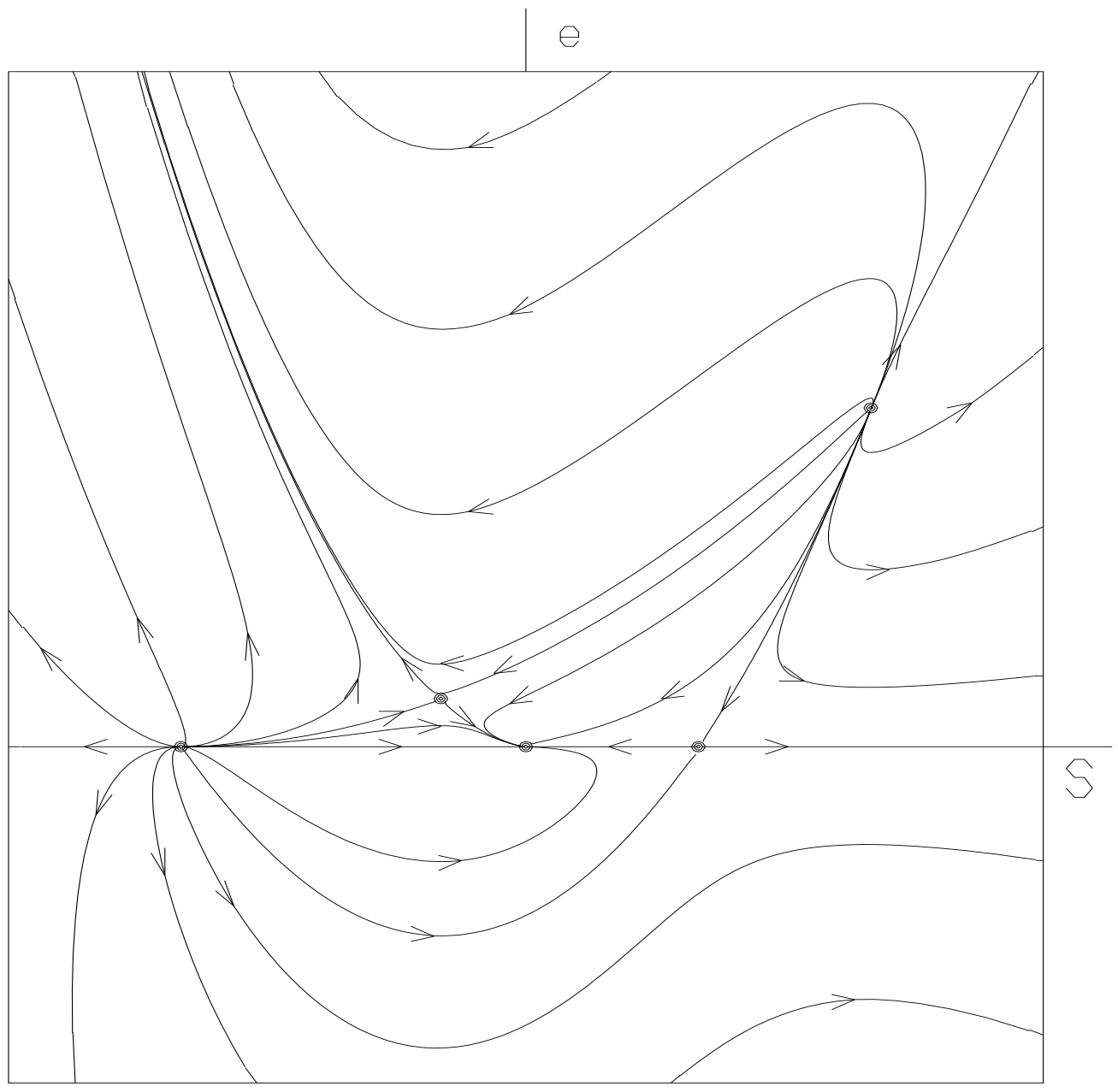
e
S

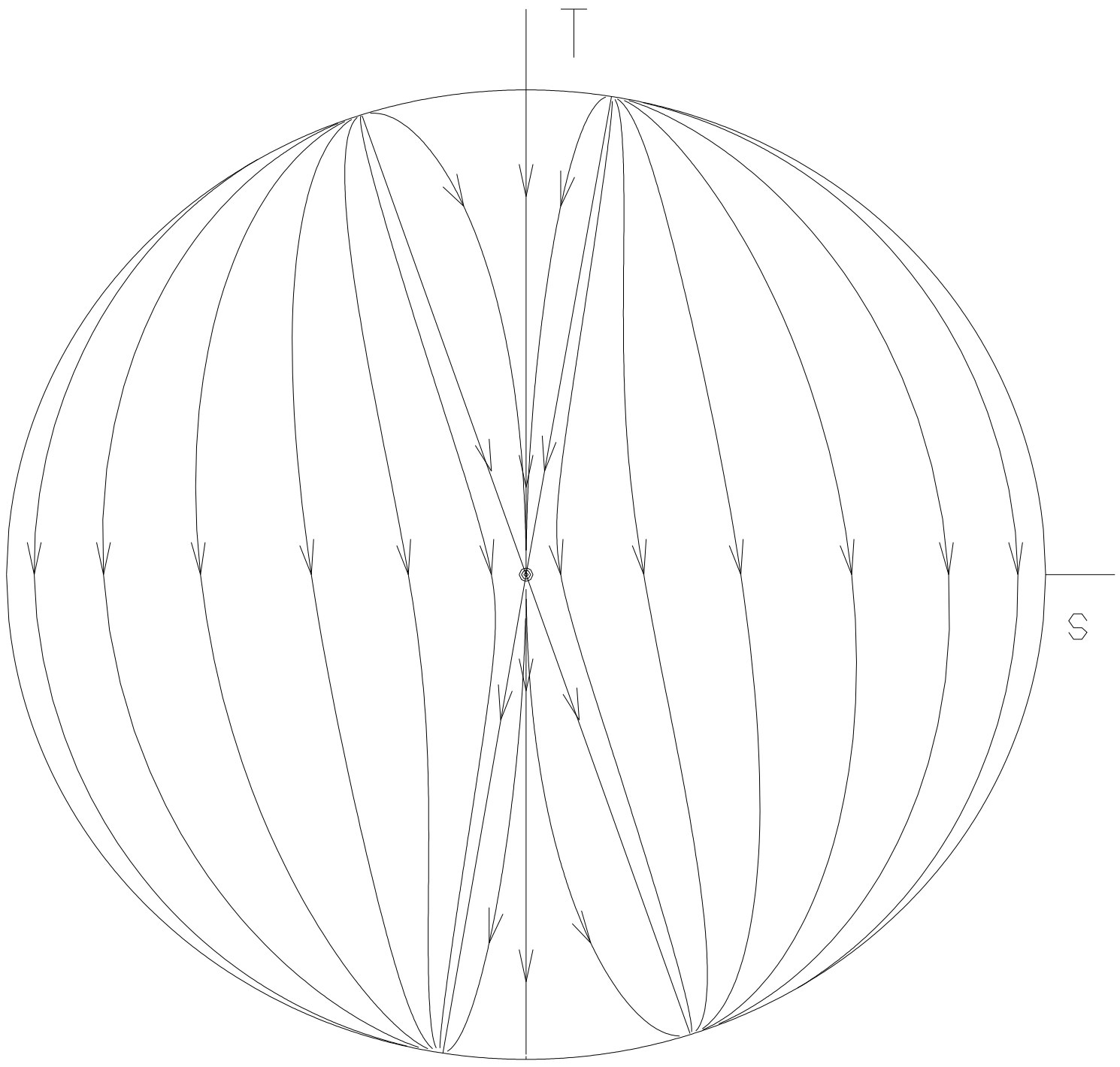

T
S

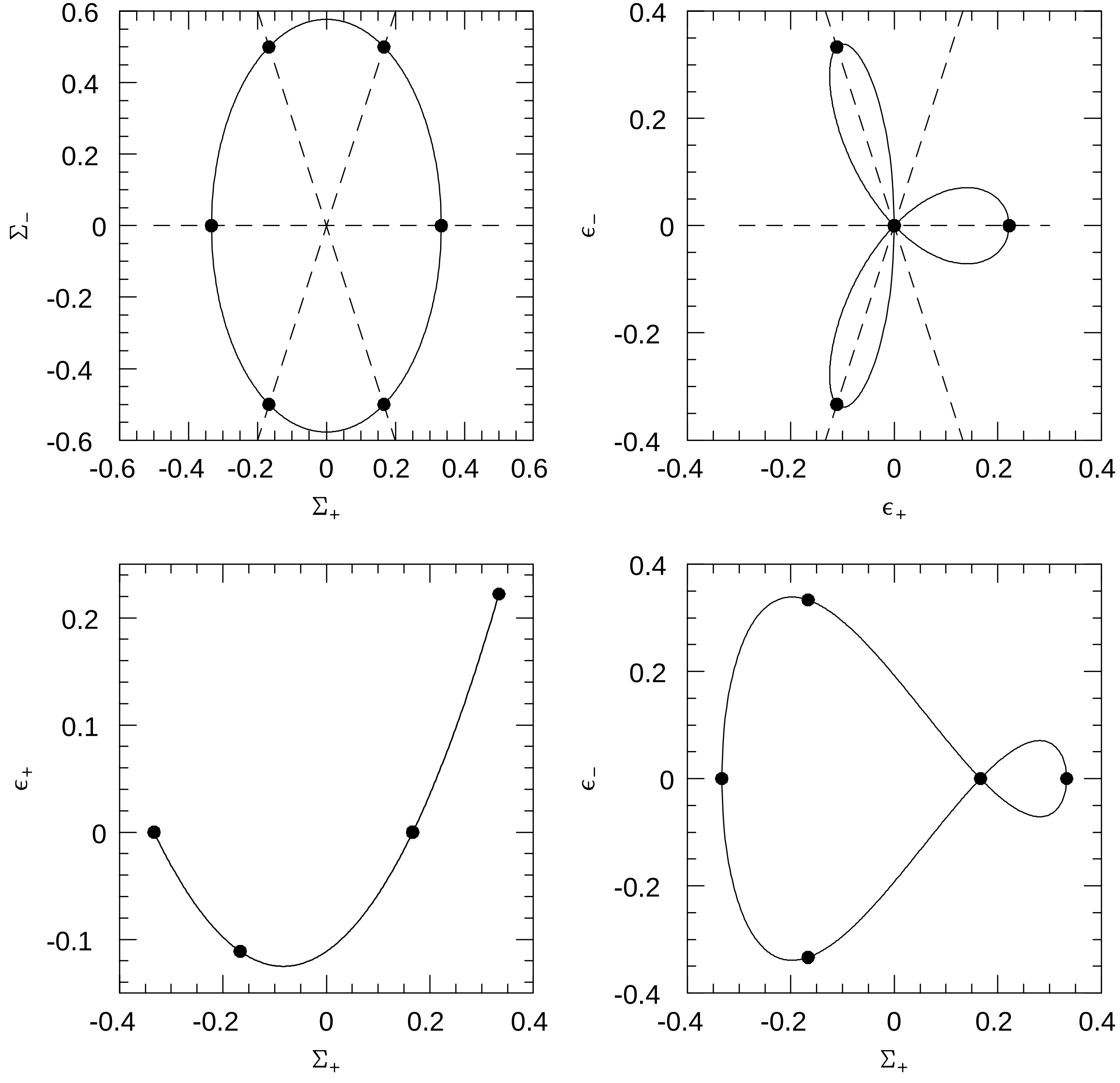